\documentclass[11pt]{article}
\usepackage{fullpage}
\pdfoutput=1
\usepackage{graphicx}
\usepackage{amsmath,amssymb,amsthm}
\usepackage{hyperref}
\usepackage{url}
\usepackage{mathtools}
\usepackage{changepage}
\usepackage{multirow}
\usepackage{subfig}
\usepackage{stackrel}
\usepackage{enumerate}

\DeclareMathOperator*{\argmax}{argmax}

\usepackage{algorithm}
\usepackage{algorithmic}

\newcommand{\floor}[1]{\left\lfloor #1 \right\rfloor}

\newcommand{\remove}[1]{}

\newenvironment{ourproof}[1][{}]{
  \begin{trivlist}\item[]\textit{Proof #1}\quad}%
  {\hfill\hspace*{\fill}~$\square$\end{trivlist}}

\theoremstyle{plain}
\newtheorem{theo}{Theorem}
\newtheorem{define}[theo]{Definition}
\newtheorem{lem}[theo]{Lemma}

\newtheorem{corr}[theo]{Corollary}
\newtheorem{ourclaim}[theo]{Claim}

\theoremstyle{definition}

\theoremstyle{remark}
\newtheorem{remark}{Remark}

\newenvironment{keyword}[1][Keywords]{\begin{trivlist}
\item[\hskip \labelsep {\bfseries #1 }]}{\end{trivlist}}
\newenvironment{class}[1][AMS Subject Classifications]{\begin{trivlist}
\item[\hskip \labelsep {\bfseries #1 }]}{\end{trivlist}}

\usepackage{color}

\title{Uniformity of point samples in metric spaces using gap ratio
\footnote{Preliminary version in the {\em Proc. of Theory and
Applications of Models of Computation - 12th Annual Conference, TAMC 2015, LNCS, volume 9076,
pages 347-358}.}}
\usepackage{fancyhdr}
\author{
Arijit Bishnu
\footnote{
ACM Unit, Indian Statistical Institute, Kolkata
}
\and
Sameer Desai
\footnotemark[2]
\and
Arijit Ghosh
\footnote{
D1: Algorithms \& Complexity, Max-Planck-Institute for Informatics, Saarbr\"ucken
}
\footnote{
Supported by the Indo-German Max Planck Center for
Computer Science (IMPECS).
}
\and
Mayank Goswami
\footnotemark[3]
\footnotemark[4]
\and
Subhabrata Paul
\footnotemark[2]
}

\date{}

\begin{document}

\maketitle

\begin{abstract}
Teramoto et al.~\cite{TeramotoAKD06} defined a new measure called the \emph{gap ratio}
that measures the uniformity
of a finite point set sampled from $\cal S$, a bounded subset of $\mathbb{R}^2$.
We generalize this definition of measure over all metric spaces by appealing
to covering and packing radius. 
The definition of gap ratio needs only a metric unlike discrepancy, a widely 
used uniformity measure, that depends on the notion of a range space and its 
volume. We also show some interesting connections of gap ratio to Delaunay
triangulation and discrepancy in the Euclidean plane. The major focus of this
work is on solving optimization related questions about selecting uniform point
samples from metric spaces; the uniformity being measured using gap ratio. We
consider discrete spaces like graph and set of points
in the Euclidean space and continuous spaces like the unit square and path
connected spaces.
We deduce lower bounds, prove hardness and approximation
hardness results. We show that a general approximation algorithm framework gives different
approximation ratios for different metric spaces based on the lower bound we deduce. Apart
from the above, we show existence of coresets for sampling uniform points from the Euclidean
space -- for both the static and the streaming case. This leads to a
$\left( 1+\epsilon \right)$-approximation algorithm for uniform sampling from the Euclidean space.


\begin{class}
52C99, 68Q25, 68R99
\end{class}

\end{abstract}

\section{Introduction}
\label{sec:intro}

Teramoto et al.~\cite{TeramotoAKD06} introduced a new measure of uniformity for point samples, 
calling it gap ratio, motivated by combinatorial approaches and applications in digital 
halftoning~\cite{Asano06,AsanoKOT02,AsanoKOT03,SadakaneTT02}. They defined the problem of minimizing 
gap ratio in a unit hypercube, in an online setting.  We attempt to generalize this definition by 
removing the online nature of the problem and extending the measure over all bounded metric spaces.

\subsection{Problem definition and hardness}

\begin{define}
Let $(\cal M, \delta)$ be a metric space and $P$ be a set of $k$ points sampled from $\cal M$. Define the minimum gap as $$r_P \coloneqq \underset{p,q \in P, p \not= q}{\min} \frac{\delta\left( p,q\right)}{2}$$ The maximum gap brings into play the interrelation between the metric space $\cal M$ and $P(\subset {\cal M})$, the set sampled from $\cal M$, and is defined as $$R_P \coloneqq \underset{q \in {\cal M}}{\sup} \delta \left( q,P\right)$$ where $\delta(q,P) \coloneqq \underset{p \in P}{\min}\,  \delta \left( q,p\right)$. \remove{is the distance of $q$ from the set $P$} The gap ratio for the point set $P$ is defined as
$$
    GR_P \coloneqq \frac{R_P}{r_P}.
$$
\label{def:gapratio}
\end{define}
In the rest of the paper, we would mostly not use the subscript $P$.

Note that the space $\cal M$, as in Definition~\ref{def:gapratio} can be both continuous and discrete.
The measure itself makes sense over unbounded spaces as well, if, we remove the finiteness of $P$ from
the definition. However, we restrict our study to finite sets only.

\remove{The \emph{maximum gap} and the \emph{minimum gap} are not actually maximum and minimum of the same quantity, so the ratio Gap ratio need not be greater than $1$. See the example in \cite{Bishnu}. Consider as an example, $\cal M$ to be two unit (diameter=$1$) balls in the Euclidean plane with centres distance $10$ apart, and $P$ to be two points, one in each ball. In this case, the gap ratio $GR=\frac{R}{r} \leq \frac{2}{9}$. It should be noted that this happens due to the disconnection in the space. In a path-connected metric space, it is easy to see that the lower bound for gap ratio is $1$ as the covering radius must be greater than the packing radius (for details refer to Lemma 3 of~\cite{Bishnu}). In a geometric sense, the maximum gap is analogous to the covering radius of $P$, and the minimum gap is analogous to the packing radius of $P$. In a uniformly distributed point set, we expect the covering to be thin and the packing to be tight. Thus the gap ratio can be a good measure of estimating uniformity of point samples.}

\begin{remark}
Let us note here that the maximum gap and the minimum gap are not the maximum
and minimum of the same parameter. The minimum gap depends only on the inter
point distances of $P$, while the maximum gap depends on the structure of the
set $P$ as well as $\cal M$. Thus the gap ratio can be less than~$1$. We will
discuss lower bounds later. 
\end{remark}

In a geometric sense, the maximum gap is analogous to the minimum radius required to cover $\cal M$ with equal sized balls (i.e., \emph{covering balls}) around each point of $P$, and the minimum gap is the maximum radius of equal sized balls around each point of $P$ having pairwise disjoint interiors (i.e., \emph{packing balls}). In a uniformly distributed point set, we expect the covering to be thin and the packing to be tight. So, we expect the maximum gap to be minimized and the minimum gap to be maximized to measure uniformity. Thus the gap ratio can be a good measure of estimating uniformity of point samples.


Using the generalized definition of gap ratio, we can pose combinatorial optimization questions where $\cal M$, for example, can be a set $S$ of $N$ points, and we would like to choose a subset $P \subset S$ of $n$ points from $S$, such that the gap ratio is minimized. Asano \cite{ipl-Asano08} in his work opened this area of research, where he asked discrepancy like questions in a discrete setting. Asano opined that the discrete version of this discrepany-like problem will make it amenable to ask combinatorial optimization related questions. We precisely do that in this paper for different metric spaces. The formal statement of the problem is as follows.

\remove{ where $\cal M$, for example, can be a set $S$ of $N$ points, and we would like to choose a subset $P \subset S$ of $n$ points from $S$, such that the gap ratio is minimized.}

\begin{define}[The gap ratio problem]
Given a metric space $\left( \mathcal{M}, \delta \right)$, an integer $k$ ($k<\left\vert {\cal M} \right\vert$ if ${\cal M}$ is finite) and a parameter $g$, find a set $P \subset \cal M$ such that $\vert P \vert = k$ and $GR_P\leq g$.
\end{define}
We next discuss a theorem that shows the gap ratio problem is even hard to
approximate for general metric spaces, and this fact forms our motivation 
to study the gap ratio problem for different metric spaces. 

\begin{theo}\label{theo:genmet}
In a general metric space, it is NP-hard  to approximate the gap ratio better than a factor of 2.
\end{theo}
\begin{ourproof}
To show this hardness, we make a reduction from independent dominating set problem, where the dominating set is also independent set. This problem is known to be NP-hard \cite{Garey:1990:CIG:574848}. The independent dominating set problem statement is as follows. Given a graph $G$ and and integer $k<|V|$ does there exist a set $D$ of size at most $k$ such that $D$ is a dominating set as well as an independent set.

Let $G=\left(V,E\right)$ and $k$ be an instance of independent domination problem. We make a weighted complete graph over $V$ such that all edges present in $G$ have weight $1$ and all other edges have weight $2$. Now the metric space $\cal M$ is given by the vertex set of the complete graph and the metric is defined by the edge weights.

We now show that $G=(V,E)$ has an independent dominating set of cardinality $k$ if and only if there exists a sampled set $P$ in $\mathcal{M}$ of $k$ points with gap ratio $1$. Let $D$ be an independent dominating set of $G$ of cardinality $k$. Let the sampled set $P$ in $\mathcal{M}$ is given by $P=D$. Now since $D$ is independent in $G$, any two points in $P$ are at a distance of $2$ in the metric space $\mathcal{M}$. Thus $r=1$. Since $D$ is a dominating set of $G$, every point in $\cal M$ has an edge of weight $1$ with some point in $P$. Thus $R=1$. Hence, the gap ratio equals $1$.

Conversely, suppose a point set $P$ of cardinality $k$ has been selected from $\cal M$ such that the gap ratio is $1$. Note that in the metric space, the value of $R$ is either $1$ or $2$ and the value of $r$ is either $\frac{1}{2}$ or $1$. Consequently, the minimum gap ratio is $1$ and the only way that can happen is if $R$ and $r$ both take value $1$.
Now, $R=1$ means the farthest point in $\cal M$ is at a distance of $1$ from the set $P$. Thus every point in $\cal M$ has an edge of weight $1$ with some point in $P$ or is in $P$. Thus $P$ forms a dominating set in $G$. Also, as $r=1$, we have the closest pair in $P$ is distance $2$ apart. Thus, the set $P$ is an independent dominating set in $G$ of cardinality $k$.
\end{ourproof}


\subsection{Previous results}\label{ssec:prev}
\subsubsection{Uniformity measures}

There are many measures of uniformity, most notably geometric discrepancy, which has already been applied in numerous areas. For the sake of brevity, we shall, from now on, say discrepancy whenever we mean geometric discrepancy.


Discrepancy measures the highest difference between the expected (by volume) sample points in a subset with the actual number of points in the subspace. Definitions vary by restricting the ``subspace" to specific geometric objects such as half planes, rectangles, axis parallel rectangles etc. When the domain is a $d$-dimensional unit cube, the lower bound on discrepancy with axis parallel rectangles (or even with axis parallel cubes) is known to be $\Omega \left( \log^{\frac{d-1}{2}} n \right)$ for a point sample of $n$ points~\cite[Theorem 6.1]{book-matousek-1999}. For a unit square a tight bound of $\Omega \left( \log n \right)$ is known~\cite[Theorem 6.2]{book-matousek-1999}. It has also been shown that the expected star-discrepancy (the axis parallel rectangles are anchored at the origin) of $N$ points chosen uniformly at random in a $d$-dimensional unit cube is $O\left( \sqrt{\frac{d}{N}} \right)$ \cite{Doerr}. With half planes, the discrepancy in the unit square is  $\Omega \left(  n^{\frac{1}{4}} \right)$~\cite[Theorem 6.9]{book-matousek-1999}. Another variant of discrepancy, called the $L_2$-discrepancy~\cite[Section 6.1]{book-matousek-1999}, is defined a bit differently. First the following function is defined, for a sample set $P$ of $n$ points, on the unit cube,
$$
    D\left( P,C_x \right) \coloneqq \left\vert n \cdot vol\left( C_x \right) -\vert  C_x \cap P \vert \right\vert,\;\; \mbox{where} \;\;
    C_x = \left[ 0,x_1 \right) \times \ldots \left[ 0,x_d \right).
$$
The $L_2$-discrepancy, of the set $P$ is
$$
    D_2\left( P , \mathcal{C}_d \right) \coloneqq \sqrt{\underset{\left[ 0,1\right]^d}{\int} D\left( P,C_x \right)^2 dx}.
$$
The lower bound for this definition of discrepancy is also  $\Omega \left( \log^{\frac{d-1}{2}} n \right)$~\cite[Section 6.1]{book-matousek-1999,Hinrichs}.

\begin{remark}
Discrepancy as a uniformity measure uses the notion of a range space (e.g.,
rectangle, circle, hyperplane, etc.) and a volume measure over the said 
range space. Different kinds of range spaces give rise to different 
discrepancy measures. On the other hand, gap ratio as a measure of uniformity, 
uses the notion of metric only. Thus, gap ratio is solely dependent on 
distances. This feature of gap ratio makes it easier to compute unlike 
discrepancy. 

\end{remark}

\subsubsection{Gap ratio}

As mentioned before the problem was originally defined in an online setting, on a unit hypercube, wherein, the gap ratio was to be minimized at every insertion. As such all previous works on gap ratio study the problem in an online setting.

Teramoto et al.~\cite{TeramotoAKD06} proved a lower
bound of $2^{{\floor{k/2}} / {(\floor{k/2}+1)}}$ for the gap ratio in the one dimensional case where $k$ points are inserted in the interval $[0,1]$ and also proposed a linear time algorithm to achieve the same. They got a gap ratio of $2$ in 2-dimension using ideas of Voronoi insertion where the new point was inserted in the centre of a maximum empty circle~\cite{Berg}. They also proposed a local search based heuristic for the problem and provided experimental results in support.

Asano~\cite{ipl-Asano08} discretized the problem and showed a gap ratio of at most 2 where $k$ integral points are inserted in the interval $[0,n]$ where $n$ is also a positive integer and $0 < k < n$. He also showed
that such a point sequence may not always exist, but a tight upper bound on the length of the sequence for
given values of $k$ and $n$ can be proved.

Zhang et al.~\cite{ipl-ZhangCCTT11} focused on the discrete version of the problem and proposed an insertion strategy that achieved a gap ratio of at most $2 \sqrt{2}$ in a bounded two dimensional grid. They also showed that no online algorithm can achieve a gap ratio strictly less than $2.5$ for a $3 \times 3$ grid.





\subsection{Our results}
We discuss the uniformity measure gap ratio for a variety of metric spaces in this paper. For ease of exposition we break our discussion into two broad groups., continuous metric spaces and discrete metric spaces. We summarize our results in Table~\ref{tab:results}.

\begin{table}[h]
\begin{center}
\caption{Our Results}\label{tab:results}
\begin{tabular}{|c|c|c|c|c|}
\hline
\multicolumn{2}{|c|}{Metric Space} & Lower Bounds & Hardness & Approximation\tabularnewline
\hline
\hline
\multicolumn{2}{|c|}{General} & none & yes & $2$-approx. hard\tabularnewline
\hline
\multirow{4}{*}{Discrete} & Graph  & \multirow{2}{*}{$\frac{2}{3}$} & \multirow{2}{*}{yes} & approx. factor: $3$\tabularnewline
 & (connected) &  &  & $\frac{3}{2}$-approx. hard\tabularnewline
\cline{2-5}
 & \multirow{2}{*}{Euclidean}   & \multirow{2}{*}{-} & \multirow{2}{*}{-} & Coreset based\tabularnewline
&  &  &  & $\left( 1+\epsilon\right)$-algorithm
\tabularnewline
\hline
\multirow{3}{*}{Continuous} & Path-Connected & $1$ & yes & approx. factor: $2$\tabularnewline
\cline{2-5}
 & Unit Square & \multirow{2}{*}{$\frac{2}{\sqrt{3}} - o \left( 1\right) $} & \multirow{2}{*}{-} & approx. factor: \tabularnewline
& in $\mathbb{R}^{2}$ &  &  & $\sqrt{3} + o \left( 1\right)$\tabularnewline
\hline
\end{tabular}

\end{center}
\end{table}

We start by discussing the motivation of the problem in Section~\ref{sec:mot}
where we show some connections with other uniformity measures.
Section~\ref{sec:cont} focuses on continuous metric spaces, while
Section~\ref{sec:disc} deals with discrete metric spaces. In continuous metric
spaces, we consider a unit square in the Euclidean plane and path connected
spaces. In discrete metric spaces, we consider connected graphs, and a set of
points in the Euclidean space. In Section~\ref{sec:cont}, we start by showing
lower bounds for path connected spaces and the unit square, we then show 
NP-hardness for general continuous spaces, and path connected spaces.  In
Section~\ref{sec:disc}, we give a lower bound for connected graphs, hardness for
graphs and $\frac{3}{2}$-approximation hardness for graphs. This shows that
sampling points uniformly is essentially a hard problem for many metric spaces.
In Section~\ref{sec:approx}, we discuss a general approximation algorithm
framework, which gives different approximation ratios for different metric
spaces. We also discuss a coreset based $\left( 1+\epsilon
\right)$-approximation algorithm for a set of points in the Euclidean space
which can also be extended to the streaming model for the same problem
setting.


\remove{In the sections that follow we shall look at lower bounds, hardness results and approximation hardness results for a few discrete and continuous metric spaces and a general approximation hardness result. We also show that the farthest point algorithm for $k$-centre by Gonzalez~\cite{gonzalez-clustering} gives a good framework for constant factor approximation algorithms and give a $\left( 1+\epsilon \right)$-approximation algorithm when the metric space is a finite set of points in $\mathbb{R}^d$ with the $l_2$ metric.}


\remove{The above shows that both continuous and discrete versions of gap ratio problem have been looked at and some efforts have been made at proving lower bounds. In this paper, we initiate a generalized study on combinatorial optimization problems related to gap ratio for different metric spaces. We also show some lower bounds, that lead to some approximation guarantees.}

\section{Motivation: connections to other measures of uniformity}
\label{sec:mot}
Generating uniformly distributed points over a bounded metric space has many applications in digital halftoning; see~\cite{Asano06,TeramotoAKD06,ipl-ZhangCCTT11} and
the references therein, numerical
integration~\cite{book-chazelle,book-matousek-1999}, computer
graphics~\cite{book-chazelle}, etc. Current techniques of
meshing with well-shaped simplices
also requires uniform distribution of points over the
region of interest~\cite[Chapter 10 \& 11]{asano-teramoto}.

For example, in the field of numerical integration uniformity of a sample in a space is measured by discrepancy. In fact, for a sequence $\left\lbrace x_1,x_2,\ldots \right\rbrace$ in an interval $\left[ 0, 1 \right]$ and any Riemann integrable complex valued function $f:\left[ 0, 1 \right] \rightarrow \mathbb{C}$, it has been proved~\cite[Chapter 1, pages 2-3]{Kuipers} 
that the sequence is uniform in $\left[ 0, 1 \right]$ if and only if
$$
    \underset{N\rightarrow \infty}{\lim} \frac{1}{N}~\stackrel[n=1]{N}{\sum}~f\left( x_n \right)
    =  \stackrel[0]{1}{\int}~f\left( x \right) dx.
$$
This result forms the basis of Monte Carlo approach to integration (although, it is not enough as the notion needs to be formalised by using random sampling and developing quantitaive theory leading to explicit upper and lower bounds on the error.) A classic result known as Koksma-Hlawka inequality says that the error in Quasi Monte Carlo integration is directly proportional to the discrepancy of point samples \cite{book-chazelle}.

\remove{ A basic application of Monte Carlo technique is as follows. Over an interval $\left[ a,b \right)$ for a sequence of $N$ points we define the Monte Carlo estimator as $ \left\langle F^N \right\rangle = (b-a) \frac{1}{N-1}~\stackrel[n=1]{N}{\sum}~f\left( x_n \right)$. A random variable $X_i \in \left[ a,b \right)$ can be generated by using a canonical random number $Z_i$, uniformly distributed in $\left[ 0,1 \right)$ as $X_i = a+ Z_i (b-a)$, and substitute this sequence in the estimator. It can be shown that the expected value of $\left\langle F^N \right\rangle$ is the desired integral and the variance is proportional to $\frac{1}{N}$. Variance reduction techniques such as importance sampling are used to refine the algorithm}

An important aspect of numerically solving partial differential equations is mesh generation. An important class of algorithms in mesh generation is \emph{Delaunay refinement algorithms}, which construct a Delaunay triangulation and refine it by inserting new vertices, chosen to eliminate skinny or oversized elements, while always maintaining the Delaunay property of the mesh~\cite[Chapter 1]{Cheng}.

\begin{figure}
\centering
\includegraphics[scale=0.90]{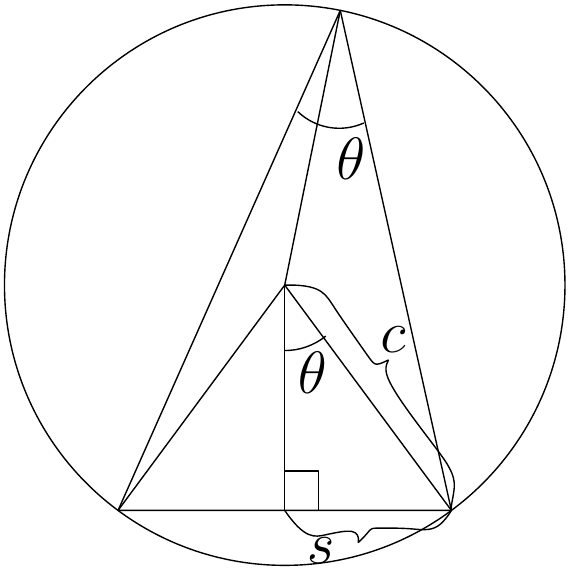}
\caption{Delaunay triangulation and gap ratio. $s \geq r$,
$c \leq R$; $\sin \theta = \frac{s}{c} \geq \frac{r}{R} =
\frac{1}{g}$.}\label{fig:Delaunay}
\end{figure}

\subsection{Connections to Delaunay triangulation}
\label{ssec:mot:del}
\begin{figure}[h]
\centering
\includegraphics[scale=1.5]{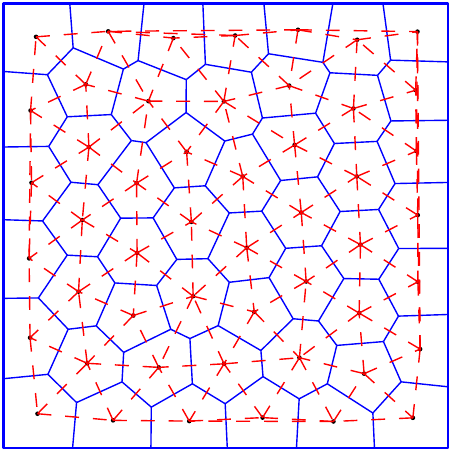}
\caption{A restricted Voronoi diagram in a square. The dashed lines are the edges of the delaunay triangulation and the solid lines inside the square form the voronoi diagram.}\label{fig:vor}
\end{figure}

We now explore the implications of gap ratio on Delaunay triangulation. Consider a set of points, $P$, in the Euclidean plane and the corresponding Delaunay graph and let $\Delta$ be the triangle in the Delaunay Graph with the smallest angle $\theta$ (Figure~\ref{fig:Delaunay}). Let us denote the maximum gap and the minimum gap corresponding to this point set by $R$ and $r$, respectively. Then the circumradius of $\Delta$ is at most $R$ and the side opposite to angle $\theta$ is at least $2r$. It is easy to see  that
$$
    \sin \theta \geq \frac{r}{R} = g^{-1}.
$$
Thus we can see that a lower gap ratio implies better Delaunay graph. The above would be true in an unbounded domain, but, as mentioned earlier we will be considering bounded metric spaces. Points of $P$ near the boundary may form triangles with circumcentre beyond the boundary of the domain and such triangles may have a circumradius which exceeds the maximum gap $R$. Voronoi diagrams~\cite{Berg} allow us to give a guarantee as to how close these triangles must be to the boundary. Consider the Voronoi diagram of $P$. For Voronoi cells which intersect the boundary of the domain we consider the Voronoi cell to be the region within the boundary (see Figure~\ref{fig:vor}). Voronoi vertices, by definition, are the points farthest from the set $P$.

A ball of radius $R$ around any point of $P$ must cover the corresponding Voronoi cell. If not, then part of the cell is being covered by a ball around some other cell, which would mean this part of the cell is closer to another site, which is not possible. Thus, each point needs to cover only the part of the domain within its corresponding cell. Triangles with circumradius greater than $R$ must have vertices with their Voronoi cell on the boundary, as the cirumcentre of such a triangle would be a Voronoi vertex beyond the boundary i.e., the vertices of the triangle must be within distance $R$ of the boundary (some of the peripheral triangles of Figure~\ref{fig:vor} are thin and have adjacent cells, their circumcentre is a Voronoi vertex of the unrestricted diagram). Thus all bad triangles of the Delaunay triangulation must be within distance $R$ of the boundary and all triangles at a distance $R$ away from the boundary must have minimum angle at least $\theta = \arcsin g^{-1}$. This lower bound also implies that no triangle has an angle greater than $180^{\circ} - 2\arcsin g^{-1}$. These bounds mean that even a gap ratio of $2$ implies that all the triangles at a distance $R$ away from the boundary have angles between $30^{\circ}$ and $120^{\circ}$, bounds guaranteed by Chew's algorithm~\cite[Page 9]{Cheng}. A lower gap ratio would give even better bounds on angles.

\subsection{An intuitive idea of uniformity}
Romero et al. \cite{Romero} suggest characterizing a uniform point sample as having the following properties,
\begin{description}
\item[Property 1] the equality with which points are spaced relative to one another in the parameter space (are they all nominally the same distance from one another?);
\item[Property 2] uniformity of point density over the entire domain of the parameter space (i.e., uniform ``coverage" of the whole domain by the set of points, and not just good uniformity within certain regions of the space); and
\item[Property 3] isotropy (no preferential directionality) in the point placement pattern.
\end{description}

\begin{remark} \label{rem:1} Note here that if the metric spaces are disconnected/discrete Properties 1 and 2 may contradict each other. For example, if our space is a subset of the Euclidean plane with the induced metric topology with two maximally connected components $A$ and $B$ such that $\underset{x\in A,y\in B}{\min} d\left( x,y\right)$ exceeds the diameter of each component, we can either satisfy Property 1 by sampling from only one component or Property 2 by sampling from both the components. Thus, this characterisation must not be considered strictly in general, but, rather it should be taken in the context of the structure of the metric space.
\end{remark}

Coming back to gap ratio as a measure, notice that the maximum gap answers the question, ``How far can we go in the metric space from the point sample?". In the case of a uniform sample one should not be able to go too far (Property 2). The minimum gap similarly answers, ``Is a pair of points too close to each other?"(Property 1). Both these quantities are independent of the axis orientation as long as the metric itself is invariant under axis orientation, thus ensuring that the gap ratio is invariant under rotation of the axes (Property 3).

Also note that since gap ratio only depends on a finite number of distances (a polynomial in the point set size) computing it is also not as difficult a task as computing measures like discrepancy.

\subsection{Measures of uniformity}
Ong et al. \cite{Ong} describe and compare three separate classes of statistical measures for the uniformity of a point sample viz., discrepancy, point-to-point measures (coefficient of variation, mesh ratio) and volumetric measures. The volumetric measures can only be used in continuous spaces. \remove{Unlike point to point measures gap-ratio sidesteps the ``contradiction" mentioned in Remark~\ref{rem:1} as the maximum gap accounts for the structure of $\cal M$ and by considering only the closest pair for minimum gap we ignore any discrepancies created by the disconnected nature of $\cal M$.}

The most commonly used measure for uniformity is discrepancy. There are many variations of discrepancy. As an example, for a sample $P$ of $n$ points in a $d$-dimensional unit hypercube we can consider the following quantity, $$ D^*\left( P \right) \coloneqq \underset{x,y \in \left[0,1\right]}{\sup} \left\vert xy -\frac{\vert  \left( \left[ 0,x \right] \times \left[ 0,y \right] \right) \cap P \vert}{n} \right\vert$$ This quantity is called the star discrepancy, where the range space is the set of axis parallel rectangles anchored at $\left( 0,0\right)$. The more uniform the point sample, the lower will be the star discrepancy. This is the general idea of discrepancy measures, to measure the deviation from the ``expected number of points" in various sizes/placements of similar objects. The above definition only considers axis parallel rectangles anchored at the origin.  Computing the above expression for a given point set takes $O\left( n^{\left(d-1\right)}\log^2 n\right)$ time and $O\left( n \right)$ space \cite{Dobkin}, which is quite high and can get higher depending on the class of objects we take the supremum over. Also, as noted by Ong et al. \cite{Ong} this measure depends on the orientation of the axis and thus cannot account for property 3 of uniformity as mentioned earlier.


\begin{figure}
\centering
\includegraphics[scale=0.3]{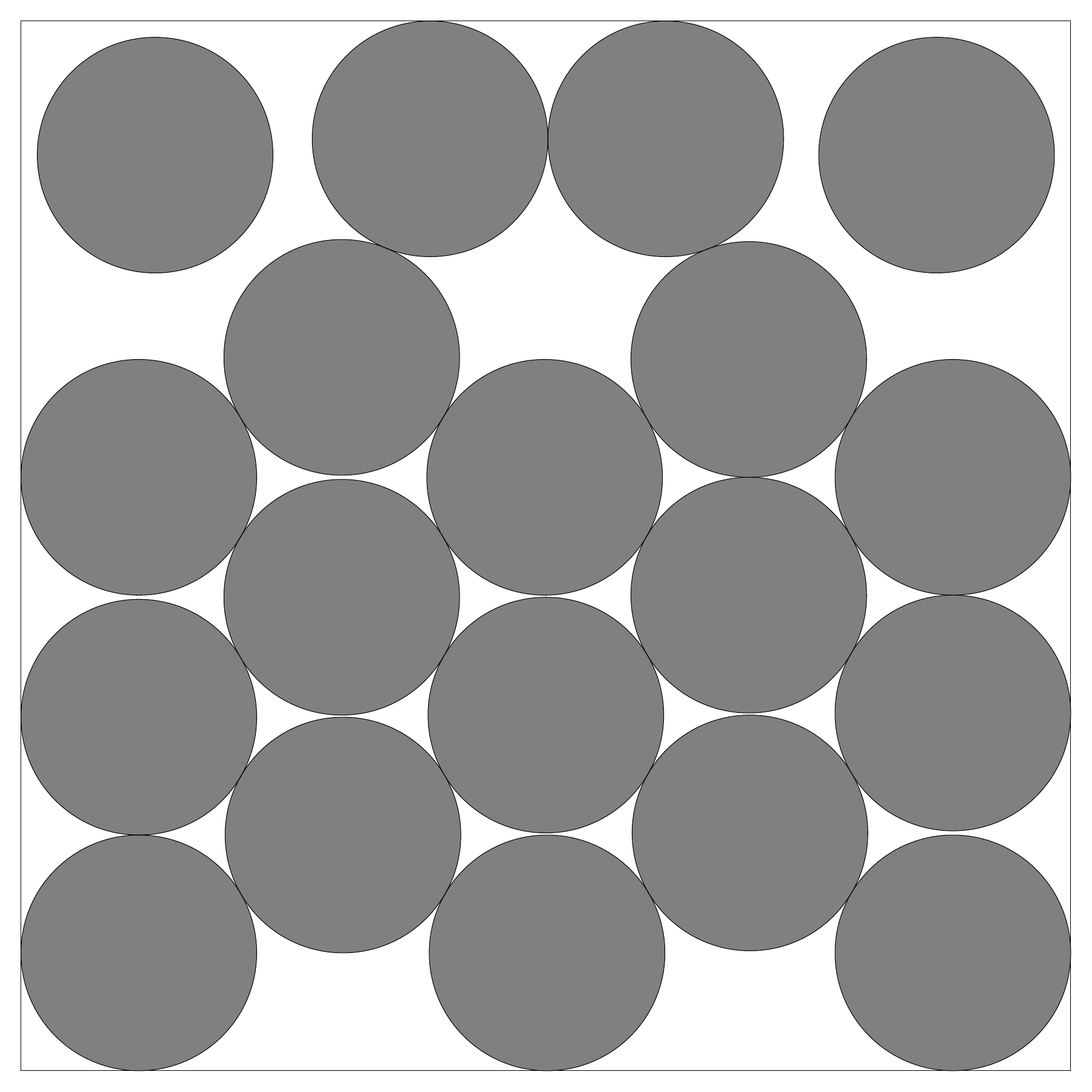}
\caption[Caption for LOF]{Optimal packing for 19 circles in the square\footnotemark }\label{fig:pacex}
\end{figure}

\footnotetext{Image provided by Dr. Eckard Specht}

Another way to ensure a uniform sample is to maximize the minimum gap. This is equivalent to packing circles of equal radius (say $r$) $\left( {\cal M} \oplus B\left( 0, r \right) \right)$ \cite{CollinsS03,Nurmela-dcg-O97,Nurmela-dcg-O99,Nurmela-cmb-O99}, where $\oplus$ denotes the Minkowski sum i.e., the set obtained by adding all pairs $(x,y)$ such that $x\in \cal M$ and $y \in B\left( 0, r \right)$ , $r$ is the minimum gap and $B\left( 0, r \right)$ is a ball of radius $r$ around the origin. Packing equal radius circles is a difficult problem~\cite{Locatelli}. This strategy does not take into account large empty areas inside $\cal M$. For example, consider the packing of 19 circles in a unit square. Figure~\ref{fig:pacex} shows the known optimal packing in such a case. Note the circles at both top corners are loose and can be moved around. However, if we attempt to minimize the maximum gap the circles would be placed in positions such that the centres are not too far from any point on the boundary of the uncovered background region.

\subsection{Relation of gap ratio to discrepancy}

We consider the star discrepancy, $D^*\left( P\right)$ for the following discussion. Consider a point set $P$ of size $n$ in the unit square. We denote the rectangle $\left[ 0,x \right] \times \left[ 0,y \right]$ by $R_{xy}$. We consider a rectangle $R_{xy}$ such that $\frac{\vert  R_{xy} \cap P \vert}{n} \geq xy$. Denote $\vert  R_{xy} \cap P \vert \coloneqq k$. Divide the rectangle into $2^l$ columns and $2^l$ rows such that $ 4^l < k \leq 4^{l+1}$. Then at least two points within the rectangle fall in one cell. So, $$r \leq \frac{(\mbox{diameter of cell)}}{2}=\frac{\sqrt{x^2 + y^2}}{2^{l+1}} < \frac{\sqrt{x^2 + y^2}}{\sqrt{k}}$$ Then
\begin{equation} \label{eq:1}
\left\vert xy -\frac{\vert  \left( \left[ 0,x \right] \times \left[ 0,y \right] \right) \cap P \vert}{n} \right\vert \leq A(x,y) \coloneqq \frac{x^2 + y^2}{r^2 n} -xy
\end{equation}

Similarly, consider another instance of $R_{xy}$ such that $\frac{\vert  R_{xy} \cap P \vert}{n} \leq xy$, and denote $\vert  R_{xy} \cap P \vert \coloneqq k$. Divide the rectangle into $2^l$ columns and $2^l$ rows such that $ 4^{l-1} \leq k < 4^{l}$. Then at least one cell will be empty. Then, $$R \geq (\mbox{diameter of cell})=\frac{\sqrt{x^2 + y^2}}{2^{l}} \geq \frac{\sqrt{x^2 + y^2}}{2\sqrt{k}}$$ Then,

\begin{equation} \label{eq:2}
\left\vert xy -\frac{\vert  \left( \left[ 0,x \right] \times \left[ 0,y \right] \right) \cap P \vert}{n} \right\vert \leq B(x,y) \coloneqq xy - \frac{x^2 + y^2}{4R^2 n}
\end{equation}

Clearly,
$$
    D^*\left( P\right) \leq \max \left\lbrace \sup_{\substack{x,y \in \left[0,1\right] \\
    \frac{\vert  R_{xy} \cap P \vert}{n} \geq xy}} A(x,y) , \sup_{\substack{x,y \in \left[0,1\right]
    \\ \frac{\vert  R_{xy} \cap P \vert}{n} \leq xy}} B(x,y) \right\rbrace.
$$
A low value for gap ratio would imply a high minimum gap and thus a low upper bound on discrepancy from inequality~\ref{eq:1} and a low maximum gap, which would give a low upper bound on discrepancy from inequality~\ref{eq:2}. Thus, we can see that a good gap ratio should give a good discrepancy as well.

Star discrepancy and discrepancy with axis parallel rectangles (where we consider all axis parallel rectangles in the square as opposed to only anchored axis parallel rectangles), $D\left( P\right)$, bound each other as $D^*\left( P\right) \leq D\left( P\right) \leq 4D^*\left( P\right)$~\cite[Chapter 2]{Kuipers}. Thus by bounding star discrepancy, we can also bound axis parallel rectangle discrepancy.

Having established connections between gap ratio vis-\`a-vis Delaunay triangulation and discrepancy, we focus on combinatorial optimization questions pertaining sampling points from metric spaces using gap ratio as the uniformity measure.

\remove{Consider $\cal M$ to be a unit square in the Euclidean plane let $P$ be a set of cardinality $k$. Note that in a plane the best way to pack circles is a hexagonal arrangement. Thus if the minimum gap is $r$ then any two points are at least distance $2r$ apart. Thus, in a rectangle of width $x$ and height $y$ we can have at most $\lfloor \frac{x}{2r} \rfloor \times 2 \lfloor \frac{\sqrt{3} y}{2r} \rfloor$ points and area of the largest empty rectangle is at most $1-\frac{2kr^2}{\sqrt{3}}$.}


\section{Continuous metric spaces}
\label{sec:cont}
\subsection{Lower bounds}
\label{ssec:lowerboundc}
Here we study the lower bounds for the gap ratio in continuous metric spaces. We first see an example, which shows us that there does not exist a general lower bound on gap ratio for a continuous metric space.

Given an $\epsilon >0$, let us consider two balls (say $A$ and $B$) in $\mathbb{R}^d$ of diameter $1$ with distance between their centres being $\frac{2}{\epsilon}+1$, where $0<\epsilon<1$. The metric space $\cal M$ is defined as $A\cup B$ and the set $P$ is defined by two points, one each in $A$ and $B$. In this case, the distance between the two points in $P$ must be at least $\frac{2}{\epsilon}$. Hence $r\geq \frac{1}{\epsilon}$ and $R\leq 1$. Thus the gap ratio becomes less or equal to $\epsilon$.


However, if the space is path connected we can fix a general lower bound.
\begin{lem}\label{lem:pathconlb}
The lower bound on gap ratio is $1$ when $\mathcal{M}$ is path connected.
\end{lem}
\begin{ourproof}
In a connected metric space $\left( \mathcal{M} , \delta \right)$, consider a sampled point set $P$. Suppose the closest pair of points $x,y\in P$ is distance $2r$ apart. Consider disks of radius $r$ around each point of $P$. This set of disks must have pairwise disjoint interiors as $x$ and $y$ are the closest pair of points in $P$. Consider a point $z\in \mathcal{M}$ on the boundary of the disk around $x$. There must be such a point, else, we have a contradiction to path-connectedness of $\mathcal{M}$. Note that $z$ must be at distance $r$ from $P$. Hence, $R\geq r$, and the lower bound follows.
\end{ourproof}

Next we consider the metric space, $\left[ 0,1 \right]^2 \subset\mathbb{R}^2$ as in Teramoto et al.'s problem~\cite{TeramotoAKD06}.  To prove the lower bound on gap ratio, we appeal to packing and covering. To find a possible lower bound on the gap ratio, we would want to increase $r$ and reduce $R$, as much as possible. To this end we will use the notions of packing and covering densities.
\begin{define}[Packing and covering densities~\cite{kuperberg,toth}]
The density of a family $\cal S$ of sets with respect to a set $C$ of finite positive Lebesgue measure is defined as $$d\left(\mathcal{S},C\right)=\frac{\underset{S\in\mathcal{S},S\cap C\neq\emptyset}{\sum\mu\left(S\right)} }{\mu\left(C\right)},$$ where $\mu$ is the Lebesgue measure. If $C$ is the plane, then we define the density as follows. Let $C\left(r \right)$ denote the disk of radius $r$ centred at the origin. Then we have $$d\left(\mathcal{S},C\right)=\underset{r\rightarrow\infty}{\lim}d \left( \mathcal{S}, C \left( r \right) \right).$$ If the limit on the right hand side does not exist, then we consider lower density defined by $$d_{-}\left(\mathcal{S}\right)=\underset{r\rightarrow\infty}{\lim}\inf d\left(\mathcal{S},C\left(r\right)\right),$$ and the upper density defined by $$d_{+}\left(\mathcal{S}\right)=\underset{r\rightarrow\infty}{\lim}\sup d\left(\mathcal{S},C\left(r\right)\right).$$ The \emph{packing density} $d_p(K)$ of a convex body $K$ is defined to be the least upper bound of the upper densities of all packings of the plane with copies of $K$, and, analogously, the \emph{covering density $d_c(K)$} of $K$ is the greatest lower bound of the lower densities of all coverings of the plane with copies of $K$.

\end{define}
\begin{lem}
The lower bound for gap ratio is $\left( \frac{2}{\sqrt{3}}- \frac{C}{\sqrt{k}} \right)$
where $C = \frac{2^{3/2}}{3^{3/4}}$, when $k$ points are sampled from $\mathcal{M} = \left[ 0,1 \right]^2$.
\label{lem:conti-lowerbound}
\end{lem}
\begin{ourproof}
Let $2r$ be the minimum pairwise distance between the point of $P$. Consider a circle of radius $r$ around each point of $P$. This forms a packing of $k$ circles of radius $r$ in a square of side length $\left( 1+2r\right)$. Suppose the density of such a packing is $d_1$. Now, we can tile the plane with such squares packed with circles. Thus we have a packing of the plane of density $d_1$. It is known that the density of the densest packing of equal circles in a plane is $\pi/ \sqrt{12}$ \cite{kuperberg}. Then obviously $d_1 \leq \pi/ \sqrt{12}$ as we have packed the plane with density $d_1$. Hence, $$d_1=k \pi r^2 / (1+2r)^2 \leq \pi/ \sqrt{12}$$ and  consequently we have, $$r\leq \left( \sqrt{k\sqrt{12}}-2 \right) ^{-1}$$
\remove{
$$d_1=\frac{k\pi r^2}{(1+2r)^2}\leq \frac{\pi}{\sqrt{12}}$$
$$\Rightarrow (k\sqrt{12}-4)r^2 -4r-1\leq 0$$
$$\Rightarrow r\leq \frac{\sqrt{k\sqrt{12}}+2}{(k\sqrt{12}-4)}=\frac{1}{\sqrt{k\sqrt{12}}-2}$$
}

On the other hand, let $R=\sup_{x \in {\cal M}} \delta(x,P)$. Clearly, circles of radius $R$ around each point of $P$ cover $\cal M$. Suppose the density of such a covering is $D_1$. Now, we can tile the plane with this unit square. Thus we have a covering of the plane with density $D_1$. It is known that the density of the thinnest covering of the plane by equal circle is $2\pi / \sqrt{27}$ \cite{kuperberg}. Then obviously $D_1 \geq 2\pi / \sqrt{27}$ as we have covered the plane with density $D_1$. Thus we have, $$D_1=k\pi R^2 / 1 \geq 2\pi / \sqrt{27}$$ giving us $$R\geq \sqrt{2} / \sqrt{k\sqrt{27}}$$
\remove{$$D_1=\frac{k\pi R^2}{1}\geq \frac{2\pi}{\sqrt{27}}$$
$$\Rightarrow R\geq \frac{\sqrt{2}}{\sqrt{k\sqrt{27}}}$$}
Hence, the gap ratio is
$$
    \frac{R}{r}\geq \left(\sqrt{k\sqrt{12}}-2\right) \frac{\sqrt{2}}{\sqrt{k\sqrt{27}}} = \frac{2}{\sqrt{3}}- \frac{C}{\sqrt{k}},
    \;\; \mbox{where} \;\; C = \frac{2^{3/2}}{3^{3/4}}.
$$
\end{ourproof}
\begin{remark}
    Teramoto et al.~\cite{TeramotoAKD06} had obtained a gap ratio of 2 in the
    online version, whereas, the lower bound for the problem is asymptotically $\frac{2}{\sqrt{3}} = 1.1547$.
\end{remark}
\subsection{Hardness}
\label{ssec:hardnesc}
\subsubsection{General NP-hardness}

In this section, we show that the gap ratio problem is hard for continuous metric space.
To show this hardness, we reduce from the problem of system of distant representatives in unit
disks~\cite{FialaKP}. We first define the problem.

\remove{
\begin{define}[$S\left(q,l\right)$-$DR$]
Given a parameter $q>0$ and a family $\mathcal{F}=\{F_i| i\in I, F_i\subseteq X\}$ of subsets of $X$, a mapping $f:I\rightarrow X$ is called a \emph{System of $q$-Distant Representatives} (shortly an $Sq$-$DR$) if \emph{(i)} $f(i)\in F_i$ for all $i\in I$ and \emph{(ii)} distance between $f(i)$ and $f(j)$ is at least $q$, for $i,j\in I$ and $i\neq j$.
When the family $\mathcal{F}$ is a set of unit diameter disks with centres that are at least $l$ distance apart, we denote the mapping by $S\left(q,l\right)$-$DR$.
\end{define}
}

\begin{define}[$ S\left(q,l\right)$-$DR$,~\cite{FialaKP}]
Given a parameter $q>0$ and a family $\mathcal{F}=\{F_i| i\in I, F_i\subseteq X\}$ of subsets of $X$, a mapping $f:I\rightarrow X$ is called a \emph{System of $q$-Distant Representatives} (shortly an $Sq$-$DR$) if
\begin{enumerate}
\item[(1)] $f(i)\in F_i$ for all $i\in I$ and
\item[(2)] distance between $f(i)$ and $f(j)$ is at least $q$, for $i,j\in I$ and $i\neq j$.
\end{enumerate}
When the family $\mathcal{F}$ is a set of unit diameter disks with centres that are at least $l$ distance apart, we denote the mapping by $S\left(q,l\right)$-$DR$.
\end{define}

Fiala et al.~\cite{FialaKP} proved that $S(1,l)$-$DR$ is NP-hard. For the general version $S\left(q,l\right)$-$DR$, we give a proof sketch using Fiala et al.'s technique. Note that for $q\leq l$, the centres of the disks suffice as our representatives. So assume that $q>l$.

Before, we go to the hardness proof we must first view the problem from a different perspective. As mentioned in \cite{FialaKP}, the problem $S\left(q,l\right)$-$DR$ is equivalent to considering disks of diameter $q+1$ around the centres of the unit disks and asking whether we can fit disjoint disks of diameter $q$, one each inside the disks of diameter $q+1$. We call the smaller disks representatives.

\subparagraph*{Instance of Planar 3-SAT} For this hardness proof, we use a reduction from planar $3$-SAT problem, which is known to be NP-hard \cite{Kratochvl1994233}. Let $\Phi$ be a CNF formula, where each variable has one positive and two negative occurrences and each clause consists of two or three literals. Let $G_{\Phi}$ be the bipartite graph of vertex set $V\cup C$, where $V$ is the variable set and $C$ is the clause set and the edge set is defined by $E = \{ xc \, | \, x \mbox{ or } \bar{x}\mbox{ occurs in clause }c\}$. By the definition of planar $3$-SAT problem, $G_\Phi$ is planar. 

\begin{figure*}[ht]
\begin{minipage}[t]{0.5\textwidth}
 \includegraphics[scale=0.4]{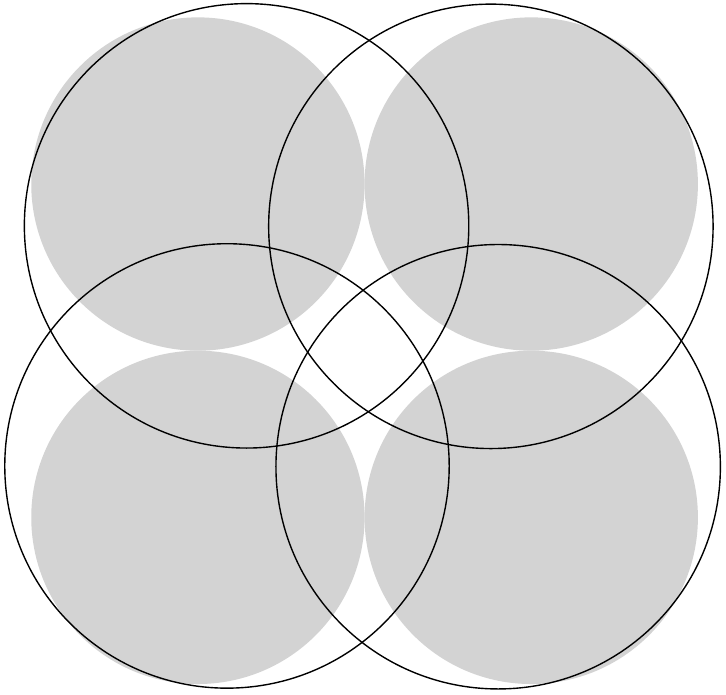}
 \centering

 {\small (a) Reserved area}
\end{minipage}
\begin{minipage}[t]{0.5\textwidth}
 \includegraphics[scale=0.4]{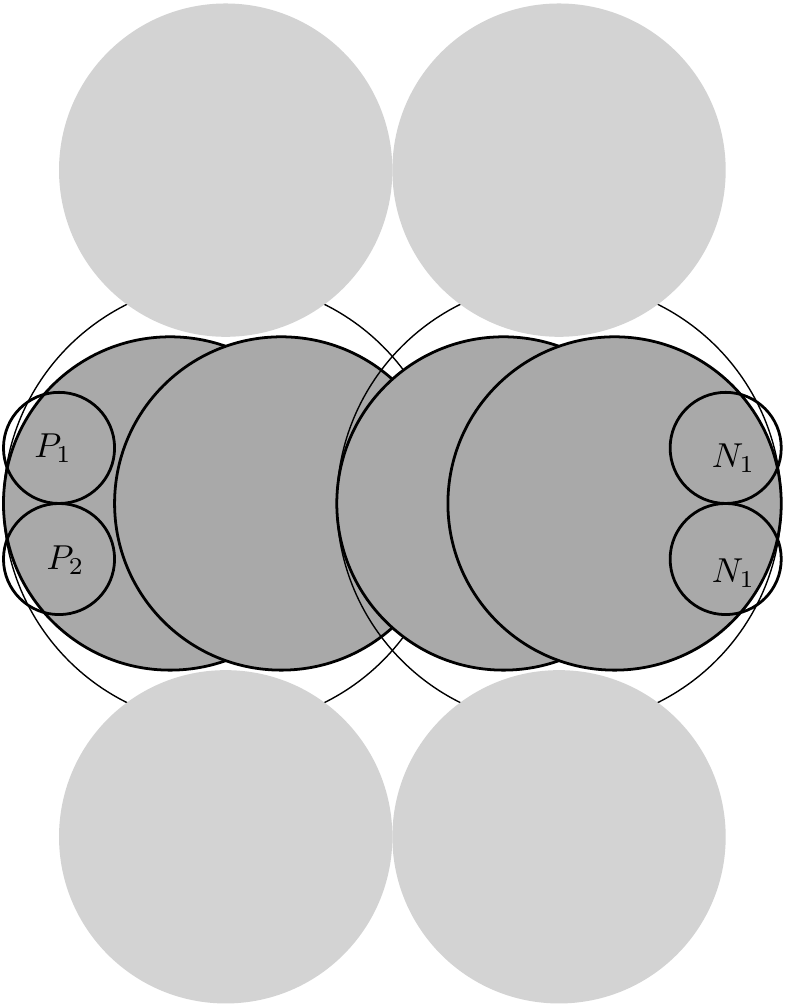}
 \centering

 {\small (b) Variable gadget}\\
\end{minipage}

\vspace*{0.2cm}

\begin{minipage}[b]{0.5\textwidth}
 \includegraphics[scale=0.4]{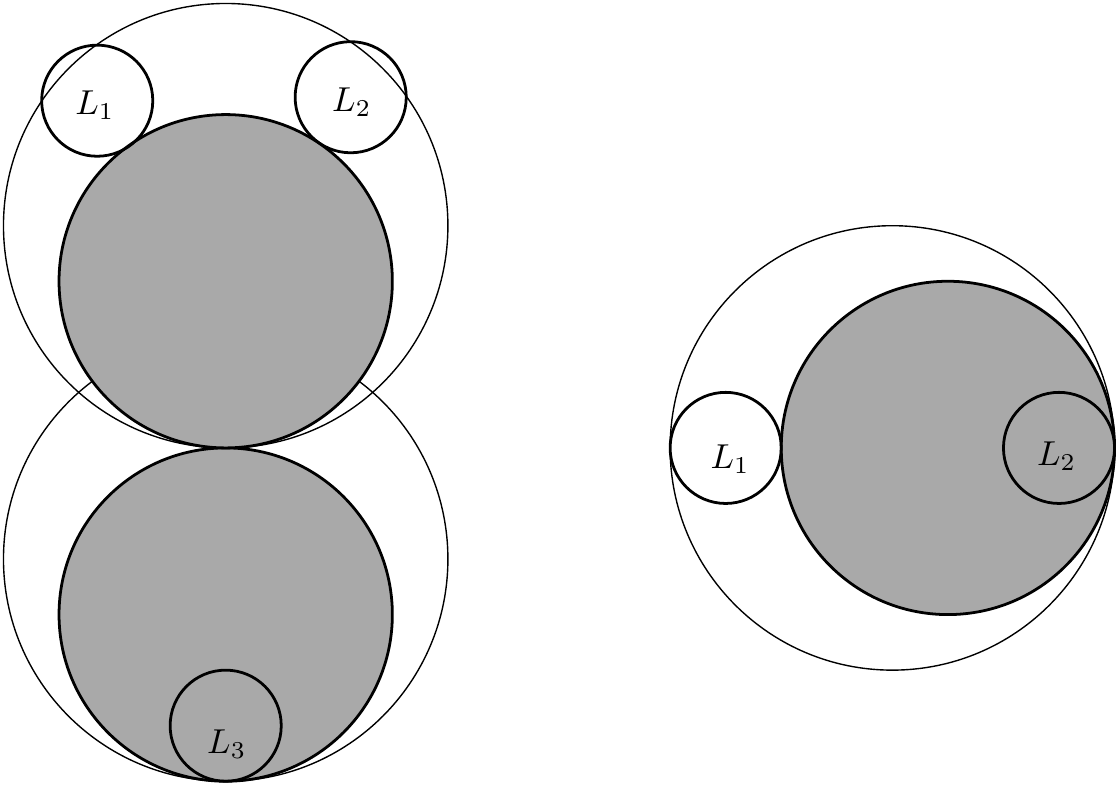}
 \centering

 {\small (c) Clause Gadget}
\end{minipage}
\begin{minipage}[b]{0.5\textwidth}
 \includegraphics[scale=0.4]{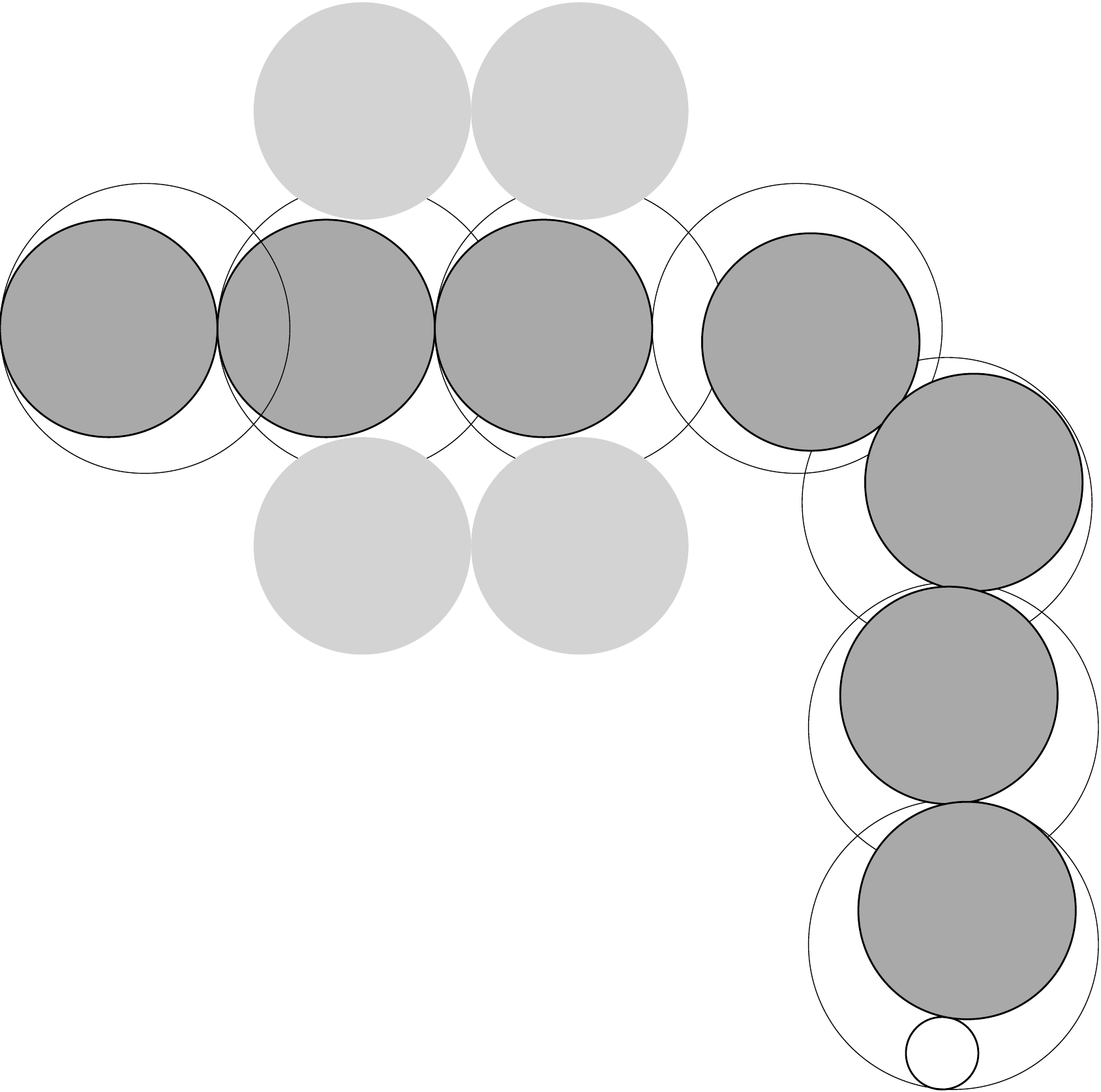}
 \centering

 {\small (d) Connector Gadget}\\
\end{minipage}

\caption{The Gadgets for reduction of planat 3-SAT to $S\left(q,l\right)$-$DR$}\label{fig:disjtunitball}
\end{figure*}

\subparagraph*{Gadgets required for reduction} Let us now discuss the gadgets, shown in Figure~\ref{fig:disjtunitball}, we use to form the corresponding $S\left(q,l\right)$-$DR$. In the figure, the small disks are auxiliary disks of unit diameter, the darkly shaded disks are sample representatives (diameter $q$) of the big disks (diameter $q+1$), which are not shaded.
\begin{description}
\item[Reserved area] The four big disks (diameter $q+1$), which are not shaded, in Figure~\ref{fig:disjtunitball}(a) allow an $S\left(q,l\right)$-$DR$ only with the four lightly shaded disks  (diameter $q$) as representatives. This arrangement forms the reserved area which is part of the variable and connector gadgets to restrict some of the big disks to having only two possible representatives. All the lightly shaded disks in Figure~\ref{fig:disjtunitball} are a part of the reserved area. For simplicity, we only show part of the reserved area (two representatives) in those gadgets.

\item[Variable Gadget] The variable gadget, shown in Figure~\ref{fig:disjtunitball}(b), consists of two big disks (diameter $q+1$), at least distance $l$ apart, and auxiliary disks corresponding to the number of occurrences of the variable. Since we are taking an instance such that each variable has one positive and two negative occurrences, we shall use gadgets with only one auxiliary disk $P1$ on one of the big disks and two auxiliary disks $N1$ and $N2$ on the other. The reserved area is placed on two sides of the gadgets so that each of the disks has only two possible representatives as shown in Figure~\ref{fig:disjtunitball}(b).

\item[Clause Gadget] The clause gadget, shown in Figure~\ref{fig:disjtunitball}(c), consists of two big disks,at least distance $l$ apart, for a clause of three literals and one big disk for a clause of two literals. There is one auxiliary disk for each literal, which are placed so that any arbitrary representatives for the big disk must intersect at least one auxiliary disk.

\item[Connector Gadget] The connector gadget, shown in Figure~\ref{fig:disjtunitball}(d), consists of a chain of big disks, at least distance $l$ apart. There is one auxiliary disk on each end of the chain, which is a part of a clause gadget or a variable gadget. We use the reserved area, as we did in the variable gadget to restrict two of the big disks to having only two possible representatives.  
\end{description}

Using the gadgets, we form the $S\left(q,l\right)$-$DR$ instance from a planar embedding of the graph $G_{\Phi}$ as follows.

\subparagraph*{The Reduction}  The variables are replaced by variable gadgets with three auxiliary disks $P_{1},N_{1}$ and $N_{2}$ such that their centres have points from $\mathbb{Z}^2$ in the direction of the edges towards the gadgets representing the clauses involving this variable (one positive and two negative occurrences). The clause gadgets are placed similarly, on the location of, and replacing, the vertices representing clauses. The edges are replaced with connector gadgets so that one of its auxiliary disks is identified by one of the variable gadgets and the other by one of the clause gadget's auxiliary disks. This forms an instance of $S\left(q,l\right)$-$DR$ problem.

We restate a generalized version of Fiala et al.'s~\cite{FialaKP} result below.

\begin{theo}\label{theo:dispersion}
$S\left(q,l\right)$-$DR$ is NP-hard for $q>l$ on the Euclidean plane.
\end{theo}
\begin{ourproof} 
Suppose, a solution of $S\left(q,l\right)$-$DR$ exists. Note that any arbitrary placement of representatives in the clause gadgets must intersect at least one of the auxiliary disks. We interpret this as the clause being satisfied by that particular literal. Note that whenever a solution of $S\left(q,l\right)$-$DR$ exists, at least one of the two auxiliary disks in the connector gadgets must intersect one of the representative disks. Also, the representative of the last disk on the other end of the connector associated to the intersected auxiliary disk $\left(L_{1},L_{2}\mbox{ or }L_{3}\right)$ must engulf the auxiliary disk of that disk (due to the reserve area limiting the kind of representatives allowed). Thus if a clause gadget representative has selected a literal with positive occurrence in it, then the auxiliary disk $P_{1}$ must be engulfed by the representative of a connector gadget. Thus, if a solution of $S\left(q,l\right)$-$DR$ exists, a representative of the disks in the variable gadget cannot intersect $P_{1}$. Now, we assign a variable $x\coloneqq \mathrm{True}$, if, in the corresponding variable gadget, the auxiliary disk $P_{1}$ is not intersected by the representatives of the variable gadget for $x$, otherwise we set $x\coloneqq \mathrm{False}$.

Conversely, given a solution of planar $3$-SAT instance, we can construct a solution of $S\left(q,l\right)$-$DR$ by using the above rule.

Hence, $S\left(q,l\right)$-$DR$ is NP-hard.
\end{ourproof}

Next we show that the above holds even for a constrained version of the problem.
\begin{lem}
$S\left(q,l\right)$-$DR$-$1$ is NP-complete for $q>l$, where $S\left(q,l\right)$-$DR$-$1$ denotes $S\left(q,l\right)$-$DR$ with one representative point constrained to lie on the boundary of one of the disks.
\end{lem}
\begin{ourproof}
Clearly, a solution to $S\left(q,l\right)$-$DR$-$1$ is a solution to $S\left(q,l\right)$-$DR$. Conversely, a solution of $S\left(q,l\right)$-$DR$ can be translated until one point hits the boundary to obtain a solution to $S\left(q,l\right)$-$DR$-$1$.

It is easy to see that $S\left(q,l\right)$-$DR$-$1$ is in NP, as any claimed solution can be checked by using a Voronoi diagram in polynomial time. Hence, it is NP-complete for $q>l$.
\end{ourproof}
We now use the above result to prove the hardness of the gap ratio problem
\begin{theo}\label{theo:gaphard}
Let $\cal M$ be a continuous metric space and $q> 2$. It is NP-hard to find a finite set $P\subset\cal M$ of cardinality $k$ such that $GR_P\leq \frac{2}{q}$.
\end{theo}
\begin{ourproof}
We show that if there is a polynomial algorithm to find a finite set $P\subset\cal M$ of cardinality $k$ such that the gap ratio of $P$ is at most $\frac{2}{q}$ for some $q> 2$, then there is also a polynomial algorithm for $S\left(q,l\right)$-$DR$-$1$.

Consider an instance of $S\left(q,l\right)$-$DR$-$1$, a family $\mathcal{F}=\left\lbrace F_1, F_2, \ldots, F_k \right\rbrace$ of $k$ disks of unit diameter such that their centres are at least distance $l$ apart, where $q>l> 2$ (even with this restriction the proof of Theorem~\ref{theo:dispersion} goes through).

We run the algorithm for the gap ratio problem $k$ times, each time on a separate instance.
The instance for the $i$-th iteration would have the disks  $\left\lbrace F_j \, \vert \, j\neq i \right\rbrace$ and a circle of unit diameter with its centre being the same as the centre of $F_i$. The following claim, whose proof follows later completes the proof.
\begin{ourclaim}\label{claim:hardball}
If a single iteration of the above process results ``yes'', then we have a solution to the $S\left(q,l\right)$-$DR$-$1$ instance.
\end{ourclaim}
\noindent Since $S\left(q,l\right)$-$DR$-$1$ is NP-hard, the gap ratio problem must also be NP-hard.
\end{ourproof}
\begin{ourproof}[of Claim~\ref{claim:hardball}]
Suppose that the gap ratio of a given point set is at most $\frac{2}{q}$ for the $i$th instance. If it so happens that two points are within the same disk, then $r\leq\frac{1}{2}$. Thus for the gap ratio to fall below $\frac{2}{q}$, we need $$R\leq\frac{2r}{q}\leq1/q<1$$ But considering the number of points that we are choosing, we must have an empty disk, which would contain a point $x$ such that $$R\geq d(P,x)\geq l- \frac{1}{2} >1$$ giving us a contradiction. Thus, we have that each disk contains exactly one point from $P$. Since, $l>2$ and $F_i$ is a circle, $R=1$. Thus, we get $r=\frac{1}{GR}\geq\frac{q}{2}$, making the closest pair to be at least a distance $q$ apart.
\end{ourproof}

\subsubsection{Path connected spaces}
\label{sssec:pathconhard}
Next, we show that it is NP-hard to find $k$ points in a path connected space such that $GR=1$. We first prove that in a path connected space it is NP-Hard to find $k$ points such that $R=r=\frac{3}{2}$. Later we extend the result for all positive real values of $r$. To this end, we need the concept of a variation of domination problem, called efficient domination problem. A subset $D\subseteq V$ is called an \emph{efficient dominating set} of $G=(V,E)$ if $|N_G[v]\cap D|=1$ for every $v\in V$, where $N_G[v]=\{v\}\cup \{x \, | \, vx\in E\}$. An efficient dominating set is also known as \emph{independent perfect dominating set} \cite{Bange19961}. Given a graph $G=(V,E)$ and a positive integer $k$, the efficient domination problem is to find an efficient dominating set of cardinality at most $k$. Note that the efficient domination problem is NP-complete~\cite{ChainChin1996147}.

\begin{figure}[h]


  \centering

  \subfloat[$\epsilon$-paths at each vertex]{\label{fig:graphtospacea}

\includegraphics[width=0.4\textwidth]{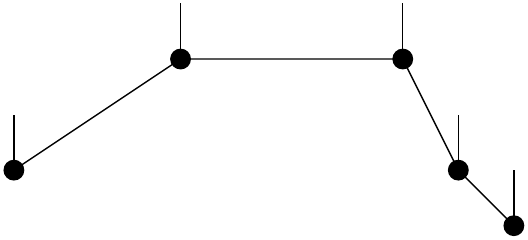}

}
\subfloat[The graph and the metric space. The open ended lines are the $\epsilon$-paths]{\label{fig:graphtospaceb}

  \centering
  \includegraphics[width=0.4\textwidth]{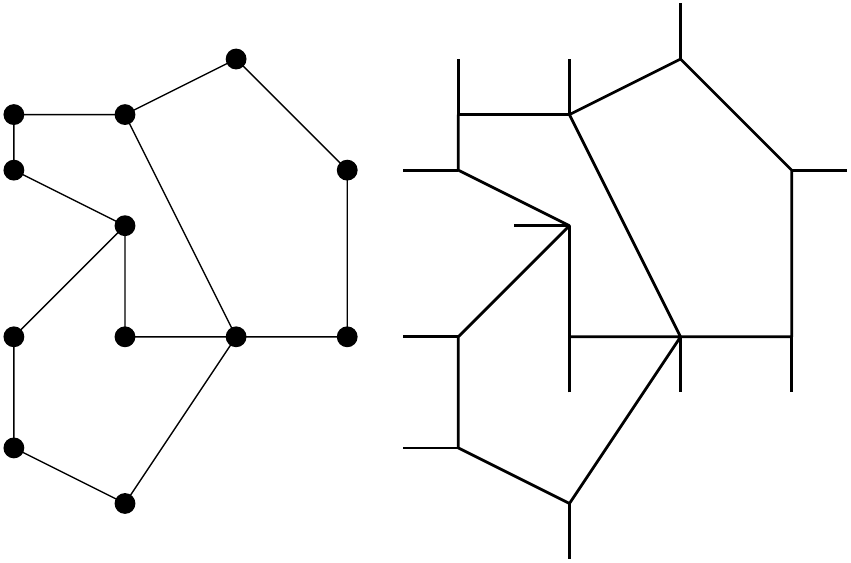}
  }\protect
 \caption{Converting a graph to a path connected space}

\end{figure}

\begin{theo}\label{theo:pathcon}
It is NP-hard to find a set $P$ of $k$ points in a path connected space $\cal M$ such that $R_P=r_P=\frac{3}{2}$.
\end{theo}

\begin{ourproof}
Let us consider an instance of the efficient domination problem, an undirected graph $G \left( V,E \right)$, and a parameter $k$. From this graph we form a metric space $\left( \mathcal{M},\delta \right)$ as follows. In $\mathcal{M}$, each edge of $E$ corresponds to a unit length path. We place at each vertex of $V$ an $\epsilon$-path, where $0< \epsilon <\frac{1}{4}$, which is merely an $\epsilon$ long curve protruding from the vertex as shown in Figure~\ref{fig:graphtospacea}. The vertices merely become points on a path formed by consecutive edges as shown in Figure~\ref{fig:graphtospaceb}. If there are edge-crossings, we do not consider the crossing to be an intersection but rather consider it as an embedding in $\mathbb{R}^3$. This ensures that different paths only intersect at vertices of the graph (this makes sure that there is direct correspondence between the path lengths in the graph and the path lengths of the metric space).  The distance, $\delta$, between two points in this space is defined by the length of the shortest curve joining the two points.

We show that finding a set $P$ of $k$ points in $\mathcal{M}$ such that $R_P=r_P=\frac{3}{2}$ is equivalent to finding an efficient dominating set of size $k$ in $G$, using a series of claims.
\begin{ourclaim}\label{claim:domtocovpack}
$D\subset V$ is an efficient dominating set in $G$ if and only if the corresponding set (i.e., each vertex of D is a point in P and vice versa) $P\subset \mathcal{M}$ has $R_P=r_P=\frac{3}{2}$.
\end{ourclaim}

Conversely, given a set $P'$ of $k$ points in $\mathcal{M}$ such that $R_{P'}=r_{P'}=\frac{3}{2}$, we want to find an efficient dominating set in $G$. If $P'\subset V$, then we are done as $P'$ is an efficient dominating set in $G$ (from the proof of Claim~\ref{claim:domtocovpack}). Otherwise, if $P'\not\subset V$, then from $P'$ we construct another set $P \subset V$ such that $R_P=r_P=\frac{3}{2}$. We form $P$ by appropriately moving points of $P'$ to the corresponding to $V$. We do this by the arguments presented below.
\begin{ourclaim}
$P'\subset V$ or $P' \cap V=\emptyset$. 
\label{claim:equalratio}
\end{ourclaim}


By Claim~\ref{claim:equalratio}, if $P'\not\subset V$, then $P'\cap V=\emptyset$. Note that in this case $P'$ cannot have midpoints of the graph edges as between any two midpoints at distance $3$ from each other, there is a vertex with an $\epsilon$-path which is distance $\frac{3}{2}$ from both points. Thus the other end of this $\epsilon$-path must be at a distance $\frac{3}{2} + \epsilon$ from both points contradicting the fact that $R_{P'}=\frac{3}{2}$. Thus each point in $P'$ must have a closest vertex. We form the set $P$ by moving each point of $P'$ to its closest vertex.
\begin{ourclaim}\label{claim:packing}
$R_P=r_P=\frac{3}{2}$.
\end{ourclaim}


By Claim~\ref{claim:packing}, without loss of generality, we can assume that the sampled set is a subset of $V$. Using ideas we present in the proof of Claim~\ref{claim:domtocovpack}, it is easy to see that, if we can find a set $P$ of $k$ points in $\mathcal{M}$ such that $R_P=r_P=\frac{3}{2}$, then we can find an efficient dominating set of $k$ vertices in $G$.

Hence, it is NP-hard to find a set $P$ of $k$ points in a path connected space such that $R_P=r_P=\frac{3}{2}$.\end{ourproof}

We now prove the claims.

\begin{ourproof}[of Claim~\ref{claim:domtocovpack}]
Let $D$ be an efficient dominating set of $G$ of cardinality $k$. Set the sampled set $P=D$. Again, note that, there cannot be a pair of vertices $x$ and $y$ in $D$ such that $\delta(x,y)<3$. This is because, if there exists a pair of vertices $x,y\in D$ with $\delta(x,y)\leq 2$, then there exists a vertex $v\in V$ such that $x,\, y\in N_G[v]\cap D$. Since, $D$ is a dominating set, if a closest pair of vertices $x,\, y\in D$ has $\delta(x,y)\geq 4$, there exists a vertex $v\in V$ such that $N_G[v]\cap D = \emptyset$. Hence, $r_P=3/2$ and balls of radius $\frac{3}{2}$ cover $\mathcal{M}$ i.e $R_P=\frac{3}{2}$.

Conversely, let $P$ be the sampled set having $k$ points with gap ratio $R_P=r_p=\frac{3}{2}$ such that P is a set corresponding to a set $D$ of vertices. Then each vertex in V is dominated by $D$ or is a vertex in $D$. And as the minimum pairwise distance is $3$ no two points dominate the same vertex (as they would be at a distance $2$ from each other. Thus $D$ must be an efficient dominating set.
\end{ourproof}

\begin{ourproof}[of Claim~\ref{claim:equalratio}]
In the proof, a path from $x$ to $y$, means the geodesic path. Suppose that $P' \cap V\neq \emptyset$ and $P' \cap V^c \neq \emptyset$. Let $x,y\in P'$ be the closest pair such that $x\in V$ and $y \notin V$. Then obviously $\delta \left( x,y \right) > 3$ as the minimum distance can be $3$ and the only points at distance $3$ from $x$ must belong to $V$. Call the first three vertices after $x$ on the path from $x$ to $y$ as $x_1,x_2$ and $x_3$. Consider a ball of radius $\frac{3}{2}$ centred at $x_2$. It must contain some point of $P'$ as $R_{P'}=\frac{3}{2}$. Let us call this point $z$. Thus $\delta \left( z,x_2 \right) \leq \frac{3}{2}$. Then $$\delta \left( z,y \right) \leq \delta \left( z,x_2 \right)+\delta \left( x_2,y \right) \leq \frac{3}{2} + \delta \left( x_2,y \right) < 2 + \left( x_2,y \right)=\delta \left( x,y \right)$$
If $z$ is a vertex then this is a clear contradiction, but if it is not a vertex then $$\delta \left( x,z \right) \leq  \delta \left( x,x_2 \right) + \delta \left( x_2,z \right) \leq 2 + \frac{3}{2} = 3.5$$ Thus $\delta \left( x,y \right) \leq 3.5$ (because $y$ must be closer than $z$ to $x$) and $\delta \left( x_2,y \right) \leq \frac{3}{2}$ as shown in Figure~\ref{fig:equa}. Consider the point at a distance $\frac{3}{2}$ from $x$ on the path from $x$ to $y$. A ball of radius $\frac{3}{2}$ centred at this point contains $x$ but not $y$ as $\delta \left( x,y \right) > 3$. Between this point and $x_2$ on the path from $x$ to $y$ there must be a point $q$ such that a ball of radius $\frac{3}{2}$ around $q$ contains neither $x$ nor $y$. Again the ball must contain at least one point $p\in P'$ (see Figure~\ref{fig:equa}) .

\begin{figure}
 \centering
 \includegraphics{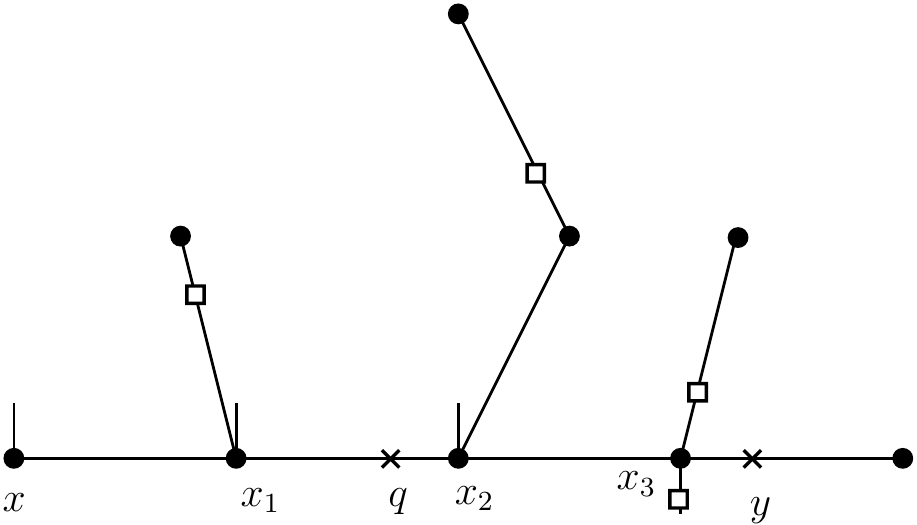}

%

 \caption{Possible positions of $p$ in proof of Claim~\ref{claim:equalratio}}
 \label{fig:equa}
\end{figure}
\begin{adjustwidth}{0.5cm}{0.5cm}
\begin{description}
\item[$p$ is on an edge of $x_1$:] Then $\delta \left( x,p \right) < 3 $ which is a contradiction.

\item[$p$ is on an edge of a neighbour of $x_2$ other than $x_1$ and $x_3$:] Then we have $$\delta \left( p,q \right) = \delta \left( p,x_2 \right) + \delta \left( x_2,q \right) \leq \frac{3}{2}$$ i.e., $\delta \left( p,x_2 \right) < \frac{3}{2}$. Thus, $$\delta \left(p,y \right) \leq \delta \left( p,x_2 \right) + \delta \left( x_2,y \right) < 3$$ which is a contradiction.

\item[$p$ is on an edge of $x_3$:] Then $\delta \left( p,y \right) < 3 $ which is a contradiction.
\end{description}
\end{adjustwidth}

\noindent Thus such a point $q$ cannot exist. Then it means such a pair $x$ and $y$ cannot exist either. Hence we have $P'\subset V$ or $P' \cap V=\emptyset$.

This proves the claim.
\end{ourproof}

\begin{ourproof}[of Claim~\ref{claim:packing}]
We have $r_{P'}=\frac{3}{2}$.

Suppose $r_P<\frac{3}{2}$. Now note that we are moving our points to the closest vertex to get $P$. Suppose we obtain the pair of vertices $u,v\in P$ from $x,y \in P'$ such that $\delta \left( u,v \right) <3$ i.e., $\delta \left( u,v \right) \leq 2$. Then $\delta \left( u,x \right) <0.5$ and $\delta \left( y,v \right) <0.5$. Thus, $$\delta \left( x,y \right) \leq \delta \left( u,x \right) + \delta \left( u,v \right) + \delta \left( y,v \right) <0.5+2+0.5=3$$ This is a contradiction.

Suppose $r_P>\frac{3}{2}$. Suppose we obtain the pair of vertices $u,v\in P$ from $x,y \in P'$ such that $\delta \left( u,v \right) >3$ i.e., $\delta \left( u,v \right) \geq 4$. Then $\delta \left( u,x \right) <0.5$ and $\delta \left( y,v \right) <0.5$.Thus, $$\delta \left( u,v \right) \leq \delta \left( u,x \right) + \delta \left( x,y \right) + \delta \left( y,v \right) <0.5+3+0.5=4$$ Again, we have contradiction.

Thus we have $r_P=\frac{3}{2}$.

We must have $R_P\geq \frac{3}{2}$ by Lemma~\ref{lem:pathconlb}.

\begin{figure*}

 \begin{minipage}[b]{0.5\textwidth}
  \centering
  \includegraphics[scale=0.80]{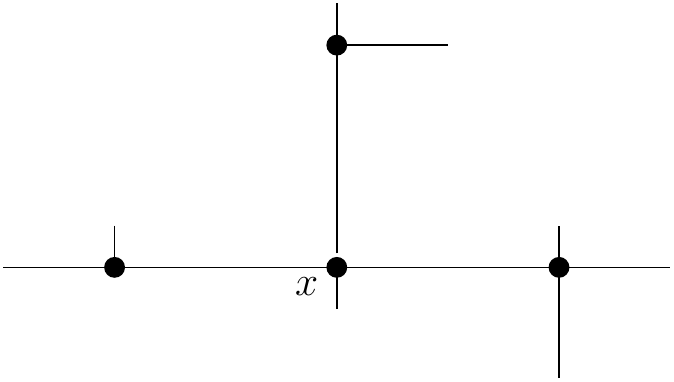}

  {\small (a)}
 \end{minipage}
 \begin{minipage}[b]{0.5\textwidth}
  \centering
  \includegraphics[scale=0.80]{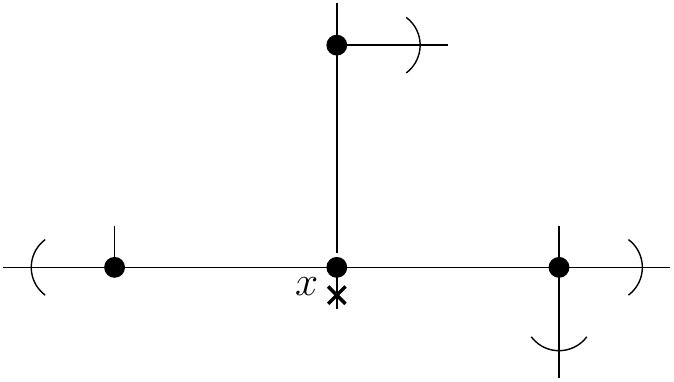}

  {\small (b)}
 \end{minipage}

 \begin{minipage}[b]{0.50\textwidth}
  \centering
  \includegraphics[scale=0.80]{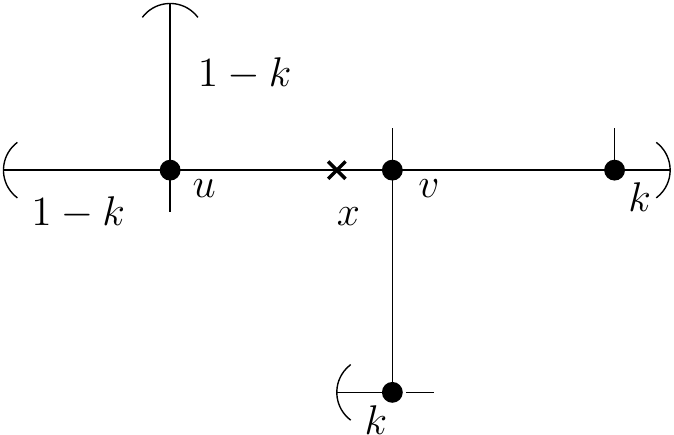}

  {\small (c)}
 \end{minipage}
 \centering

 \caption{(a) Case 1: $x$ is a vertex, (b) Case 2: $x$ is on an $\epsilon$-path. We have taken the ball from case 1 and the brackets denote the boundary of the ball in this case, (c) Case 3: $x$ is on a full edge}
 \label{fig:ball}
\end{figure*}


Suppose we have $R_P>\frac{3}{2}$. Then there is a point $x$ such that a ball of radius $\frac{3}{2}$ doesn't contain any point of $P$. But it must contain a point of $P'$. Again let us consider cases.

\begin{adjustwidth}{0.5cm}{0.5cm}
\begin{description}
\item[1) $x$ is a vertex:] Then the ball around $x$ must contain only full edges and edges of length half (see Figure~\ref{fig:ball}(a)). It also contains a point (say $y$) of $P'$. The closest vertex of any such point must be inside this ball. This gives us a contradiction.

\item[2) $x$ is on an $\epsilon$-path]: This case is similar to the previous case as the ball in this case would clearly be a subset of the ball in the previous case (see Figure~\ref{fig:ball}(b)).

\item[3) $x$ is on a full edge]: In this case it is important to note that every point in the ball around $x$ lies on a path (geodesic) that goes through $x$ and lies completely within the ball. Each path is of length $3$ or less and $x$ is at the centre of the path (see Figure\ref{fig:ball}(c)). Let $v$ be the nearest vertex of $x$ and $u$ be the other vertex of the edge on which $x$ lies. Thus all neighbours of $v$ are in the ball.Thus each path of length $3$ inside the ball is formed by  one edge of length $l$ (where $l = \frac{1}{2} - \delta \left( x,v \right)$), two edges of length $1$ (including $uv$) and one edge of length $1-l$ (see Figure~\ref{fig:ball}). Note that edges of length $1-l$ are incident on $u$ and edges of length $l$ are incident on neighbours $v$ (excluding $u$).Let us say a point $y\in P'$ lies in this ball (without loss of generality we may assume that $y$ is on a full edge as an $\epsilon$-path cannot intersect with the ball without the corresponding vertex being in the ball). Thus if $y$ lies on one of the edges of length $1$ its closest vertex will be $u$, $v$ or a neighbour of $v$ all of which are in the ball. If $y$ lies on one of the edges of length $l$ then its closest vertex will be a neighbour of $v$ which is in the ball. So assume $y$ lies on an the edge of length $1-l$. Thus if $y$ is within distance $\frac{1}{2}$ of $u$ then the closest vertex for $y$ is $u$ which is also a contradiction. Thus let us assume that $y$ is more than distance $\frac{1}{2}$ of $u$.

%


Then there is a point $w$ between $x$ and $v$ such that a ball of radius $\frac{3}{2}$ has $y$ on its boundary. Again this ball will contain paths of length at most $3$. And the paths of length $3$ can be characterized by one edge of length $l_1$ ($l_1 = \frac{1}{2} - \delta \left( w,v \right)$), two edges of length $1$ (including $uv$) and one edge of length $1-l_1$. The edges of length $1-l_1$ are subsets of the edges of length $1-l$ (the difference is $\delta \left( w,x \right)$ ). Now if the only point in $P'$ on the boundary of this ball is $y$ then between $w$ and $v$ we must have a point such that a ball of radius $\frac{3}{2}$ centred around it does not intersect $P'$ at all which is not possible. Hence there must be another point $p\in P'$ at distance of $3$ from $y$ such that $x,w$ and $v$ are on the path from $y$ to $p$ (because $\delta \left( w,p \right) = \frac{3}{2}$ and as mentioned earlier there are only two such kind of points and if $p$ is on an edge of length $1-l_1$ then $\delta \left( y,p \right) = 2\left(1 - l_1 \right) < 3$). Then $p$ is on an edge of length $l_1$ in which case the closest vertex to $p$ is a neighbour of $v$ which was in the ball around $x$. Thus again we have a contradiction.



\end{description}
\end{adjustwidth}

This proves the claim.
\end{ourproof}

In the above reduction, taking the edge lengths to be $\frac{2x}{3}$ instead of $1$ and $\frac{2x\epsilon}{3}$-paths instead of $\epsilon$-paths we have that it is NP-hard to find a set of $k$ points in a path connected space such that $R_P=r_P=\frac{3}{2} \times \frac{2x}{3} = x$. Since this can be done for any positive $x$, the following theorem follows as a corollary to Theorem~\ref{theo:pathcon}.
\begin{theo}
It is NP-hard to find a set of $k$ points in a path connected space such that gap ratio is $1$.
\end{theo}

\section{Discrete Metric Space}
\label{sec:disc}
\subsection{Graph}
\subsubsection{Lower Bounds}
\label{sssec:lowerboundd}
Here we study the lower bounds for the gap ratio problem
in discrete metric spaces. We start by giving an example which demonstrates the lack of a general lower bound for discrete metric spaces. Given any $\epsilon >0$, we construct an example of a discrete metric space and a sampled set admitting a gap ratio $\epsilon$.

\begin{figure}
 \centering
 \includegraphics[scale=.7]{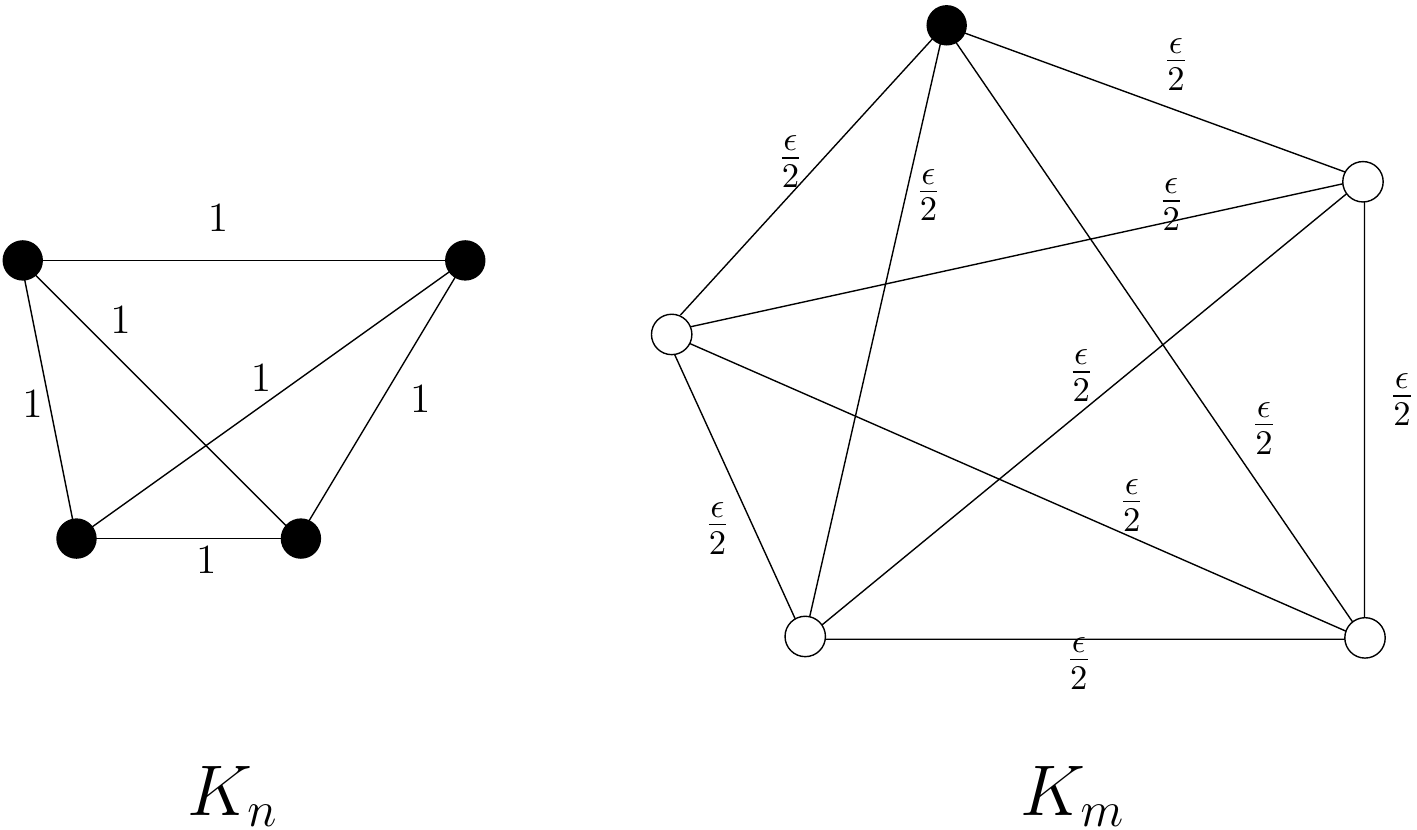}
 \caption{Lower bound for the discrete case. Filled in vertices form the set $P$. ${\cal M} = V\left[K_{n}\right]\cup V\left[K_{m}\right]$}
\end{figure}

Consider the complete graph $K_{n}$ for some $n\in\mathbb{N}$ with each edge
having unit weight and the complete graph $K_{m}$ for some $m\in\mathbb{N}$ with
each edge having weight $\frac{\epsilon}{2}$. Let $V[G]$ denote the vertex set
of the graph $G$. Now suppose that the metric space $\left({\cal
M},\delta \right)$ is $V\left[K_{n}\right]\cup V\left[K_{m}\right]$
with the metric $\delta$ being the edge weights when there are edges between
vertices of $V\left[K_{n}\right]\cup V\left[K_{m}\right]$ and
$\infty$, otherwise. Let the sampled set $P=V\left[K_{n}\right]\cup\left\{ v\right\} $ for some $v\in V\left[K_{m}\right]$. We have $\mbox{min}_{q \in P} \delta(p,q) = \frac{\epsilon}{2}$ for all
$p\in{\cal M}\setminus P$. Thus, $R=\frac{\epsilon}{2}$. By the definition
of $P$, $r=\frac{1}{2}$. Thus, $GR=\frac{R}{r}=\epsilon$.

First we prove the lower bound of gap ratio on a metric space $\cal M$ which is the vertex set $V$ of an undirected connected graph $G=(V,E)$. The distance between a pair of vertices is the length of the shortest path between them.
\begin{lem}\label{lem:lbgraph}
Gap ratio has a lower bound of $\frac{2}{3}$ when the metric space
$\cal M$ is a connected undirected graph. The bound is achieved only when $R=1$ and $r=\frac{3}{2}$.
\end{lem}
\begin{ourproof}
Suppose a set of vertices $P\subset \cal M$ is sampled. Let a closest pair of vertices in $P$ be distance $q$ apart. Thus, $r=\frac{q}{2}$. Now between these two vertices, there is a path of $q-1$ vertices in $\mathcal{M} \setminus P$. Among these $q-1$ vertices, the vertex farthest from $P$ is at a distance $\left\lfloor \frac{q}{2} \right\rfloor$ from $P$. Thus, $R\geq \left\lfloor \frac{q}{2} \right\rfloor$ and $$GR = \frac{R}{r}\geq \frac{2}{q} \left\lfloor \frac{q}{2} \right\rfloor$$ Note that, when $q=1$, clearly we have a gap ratio greater or equal to $2$. Now, we analyse this expression for even and odd values of $q$. If $q$ is even,  $$GR\geq \frac{2}{q} \left\lfloor \frac{q}{2} \right\rfloor = \frac{2}{q} \frac{q}{2}=1$$ and if $q$ is odd and $q\geq 3$, $$GR\geq \frac{q-1}{q}$$ Since this function is monotonically increasing, $GR \geq \frac{2}{3}$, and the equality only occurs for $q=3$.

Thus, the gap ratio $GR=\frac{2}{3}$ implies $q=3$, which means $r=\frac{3}{2}$. Therefore, $R=GR\times r= 1$. Hence, $GR=\frac{2}{3}$ only when $R=1$ and $r=\frac{3}{2}$.
\end{ourproof}
\subsubsection{Hardness}
\label{sssec:hardnessd}

In this section, we show that the problem of finding minimum gap ratio is NP-complete even for graph metric space.
\begin{theo}
In graph metric space, gap ratio problem is NP-complete.
\end{theo}
\begin{ourproof}
First note that, the gap ratio problem in graph metric space is in NP. To prove the hardness, we use a reduction from efficient domination problem, to the gap ratio problem. Given an instance of efficient domination problem $G=(V, E)$ and $k$, set $\mathcal{M} = V$ as the metric space and the shortest path distance between two vertices as the metric $\delta$. Claim~\ref{claim:domtocovpack} proves the theorem.
%

\end{ourproof}

\subsubsection{Approximation Hardness}
\label{sssec:APX-Hardness}

Here we use the hardness of path connected space from Section~\ref{sssec:pathconhard} to show that the gap-ratio problem is APX-hard on the graph metric.
\begin{figure}
\centering
  \includegraphics[width=0.850\textwidth]{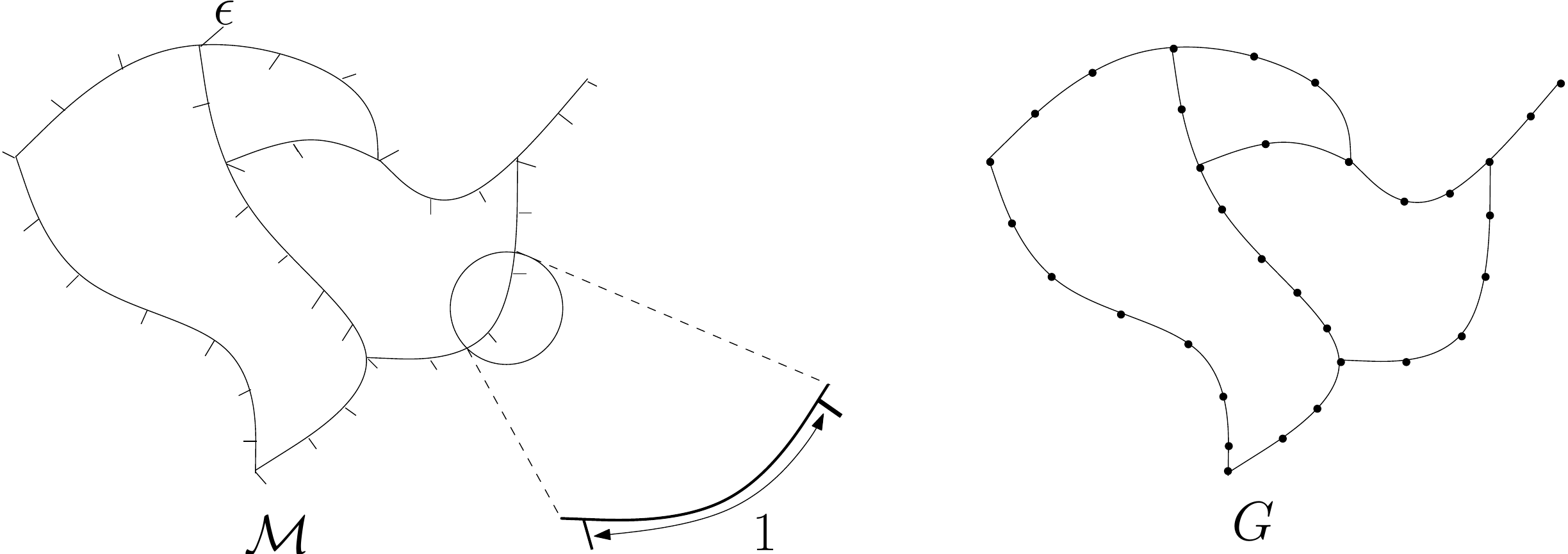}
  \caption{Illustration of the reduction in Theorem~\ref{theo:graphapprox}}\label{fig:path}
\end{figure}
\begin{theo}\label{theo:graphapprox}
In an unweighted  graph, it is NP-hard  to approximate the gap ratio better than a factor of $\frac{3}{2}$.
\end{theo}

\begin{ourproof}
In Section~\ref{ssec:hardnesc}, we reduced the problem of finding a set of $k$ points in a graph such that the gap ratio is $\frac{2}{3}$ to the problem of finding a set of $k$ points in a path-connected space such that the gap ratio is $1$. We use this hardness of gap ratio being $1$ on instances similar to the one created in the reduction to prove $\frac{3}{2}$ approximation hardness on graphs.

Our starting instance is a space formed by joining integer length curves at their ends (so that points that divide these curves into unit length curves form a connected graph with the unit length curves as edges). Also for some $0 < \epsilon < \frac{1}{4}$ we join curves of length $\epsilon$ (at one end) at points such that the integer length curves are divided into unit length curves. Let us call this path connected space $\cal M$. Note that $\cal M$ is similar to the path connected space formed in Section~\ref{ssec:hardnesc}, but, the general shape of the space may vary. The reduction is illustrated in Figure~\ref{fig:path}.  The metric on this space is defined by the length of the shortest path between pairs of points. We form the graph $G= \left( V, E \right)$ by putting vertices at the place where the $\epsilon$-length curves are joined to the integer length curves. The $\epsilon$-length protrusions are discarded and the unit length curves between the vertices form the edge set.
\begin{ourclaim}\label{claim:approxgraph}
There exists a polynomial time algorithm to find $P \subset \cal M$ such that $\vert P \vert = k$ and $R_P = r_P = \frac{2t+1}{2}$ for some $t \in \left\lbrace 1,2,..., \right\rbrace$ if and only if  there exists a polynomial time algorithm to find a set of $k$ vertices in $G$ such that the gap ratio of the set is strictly less than $1$.
\end{ourclaim}


%

This gives us that it is NP-hard to find a set with gap ratio less than $1$ in graphs, i.e it is NP-hard to find an algorithm which approximates gap ratio within a factor better than $\frac{3}{2}$.

Note here that if we could have proven Claim \ref{claim:approxgraph} for $\vert P \vert = k$ and $R_P = r_P = \frac{t}{2}$ for some $t \in \left\lbrace 2,3,..., \right\rbrace$, then we wouldn't need to say strictly less than $1$ in the statement.
\end{ourproof}

We now prove Claim~\ref{claim:approxgraph}.

\begin{ourproof}[of Claim~\ref{claim:approxgraph}]
Suppose we have a set of $k$ vertices in $G$ with gap ratio strictly less than $1$. Let $q$ be the minimum distance of a pair of points in this set. Then by proof of Lemma \ref{lem:lbgraph}, we have gap ratio is at least $\frac{2}{q}\lfloor \frac{q}{2} \rfloor$ and $r = \frac{q}{2}$. Now unless $R=\lfloor \frac{q}{2} \rfloor$, we have gap ratio greater than $1$. If $q$ is even, then the gap ratio will be at least $1$. Hence, $q$ must be odd. Thus, the corresponding point set (viz., $P$) in $\cal M$ has $R_P=r_P=\frac{q}{2}$.

Conversely, let $P \subset \cal M$, such that $\vert P \vert = k$, and $R_P~=~r_P~=~\frac{2t+1}{2}$ for some $k \in \left\lbrace 1,2,..., \right\rbrace$. Then using the ideas in Theorem~\ref{theo:pathcon} one can verify that the points can be shifted to vertices the graph vertices to get a gap ratio of $\frac{2t}{2t+1} < 1$ in the $G$.
\end{ourproof}

\section{Approximation Algorithms}
\label{sec:approx}

In this section we give a general approximation scheme and a $\left( 1+\epsilon \right)$ approximation algorithm for the case whem $\cal M$ is a finite set of points in the Euclidean metric space.

\subsection{Farthest point algorithm}\label{ssec:fpi}
Gonzalez's~\cite{gonzalez-clustering}  {\em farthest point insertion method} (with a slightly tweaked initiation) for $k$-centre (Algorithm~\ref{algo:fpi}) seems a natural generalisation of Teramoto et al.'s~\cite{TeramotoAKD06} Voronoi insertion method. Indeed it gives an upper bound of $2$ over any metric space. The following is an outline of the algorithm.

\begin{algorithm}
  \caption{Pseudocode of {\em Farthest-point-insertion$\left( \mathcal{M}, k \right)$}}
  \begin{algorithmic}[1]\label{algo:fpi}
    \STATE{\bf Input:} metric space $(\mathcal{M}, \delta)$ and $k$; // $\mathcal{M} = \{p_{1}, \, \dots, \, p_{n}\}$
    \STATE{\bf Initialize:} find
    $q_{1}, \, q_{2} \in \mathcal{M}$ with $\delta(q_{1}, q_{2}) = diam (\mathcal{M})$
    and $S_{2} = \{ q_{1}, \, q_{2}\}$;
    \FOR{$i = 2$ to $k-1$}
        \STATE $q_{i+1} \gets \argmax_{p_{j} \in \mathcal{M}} \delta(p_{j}, S_{i})$; // $q_{i+1}$ is the point farthest from $S_{i}$ in $\mathcal{M}$
        \STATE $S_{i+1} \gets S_{i} \cup \{ q_{i+1}\}$;
    \ENDFOR
    \STATE{\bf Output:} $S_{k}$ and $GR_{S_{k}} = \frac{R_{S_{k}}}{r_{S_{k}}}$;
  \end{algorithmic}
\end{algorithm}

We now analyse the algorithm. Without loss of generality, let $P = \{ p_{1}, \, \dots, \, p_{k}\}$ be the set with optimal gap ratio, and let $GR = \alpha$.

\begin{lem}\label{gapratio2}
In Algorithm 1, $R_{S_{i}}\leq R_{S_{i-1}}$ for each $i\in \{2, \ldots, \, k\}$ and the gap ratio $GR_{S_{i}}$ is at most $2$ after each iteration.
\end{lem}
\begin{ourproof}
We prove this by induction on $i$. Clearly $R_{S_2}\leq R_{S_1}$ as that is how $q_2$ is chosen.
Suppose for some $t$,
$R_{S_{j}}\leq R_{S_{j-1}}$ for $j=2,\, \dots,\, t$. Now by our scheme, $q_{t+1}$ is chosen at a
distance $R_t$ from $q_t$.
Thus, there exists a point $x$ at a distance $R_{S_{t+1}}$ from $q_{t+1}$. Hence, by definition
of $R_{S_i}$, we have $$ R_{S_{t+1}}= \delta \left(x,q_{{t+1}}\right) \leq \delta \left(x,S_{t}\right) \leq R_{S_{t}}$$

Note that at each insertion, we have chosen $q_{i}$ at a distance of $R_{S_{i-1}}$ from
$S_{i-1}$. So, $r_{S_{i}}=\frac{R_{S_{i-1}}}{2}$.
Thus, for all $i \in \{2, \, \dots, \, k\}$, we get
$$
  GR_{S_{i}} = \frac{R_{S_{i}}}{r_{S_{i}}} =2\times\frac{R_{S_{i}}}{R_{S_{i-1}}}\leq 2.
$$
\end{ourproof}
\begin{remark}
It is important to note here that when we mention discrete settings we mean the space $\cal M$ itself is discrete. Zhang et al.'s~\cite{ipl-ZhangCCTT11} result on bounded two dimensional grids (especially a $3 \times 3$ grid) computes the maximum gap over the continuous space of the square. Hence, Zhang's result that no online algorithm can achieve a gap ratio strictly less than $2.5$ for a $3 \times 3$ grid, does not contradict this lemma.
\end{remark}
 \begin{figure}
 \centering
 \includegraphics[scale=0.60]{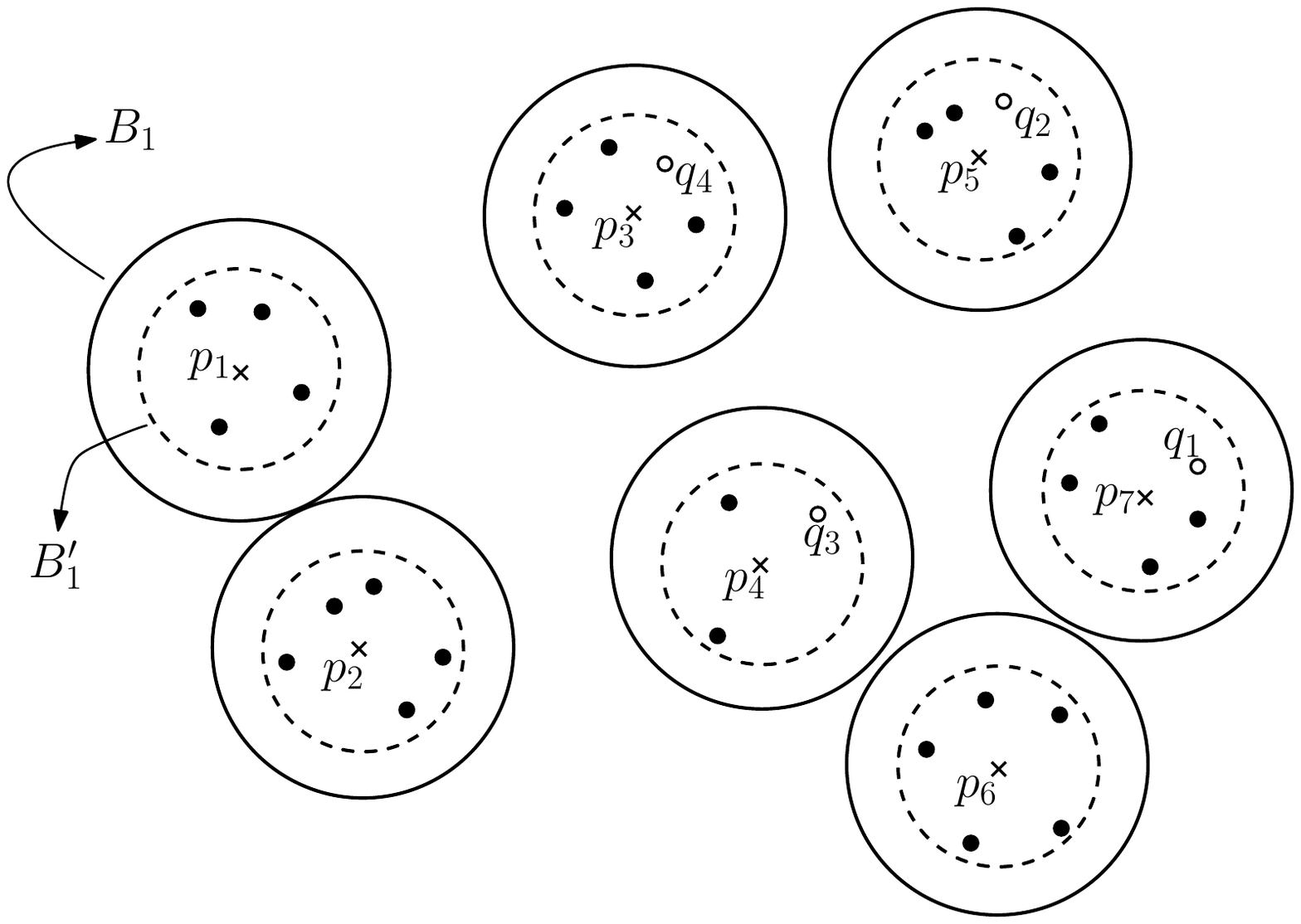}
 \caption{The cross points denote the set $P$ and the hollow points denotes the set $S_{i}$}
\end{figure}

\begin{theo}\label{thm-FPI}
 Farthest point insertion gives the following approximation guarantees:

\begin{enumerate}[(i)]

\item if $\alpha \geq 1$, then the approximation ratio is $\frac{2}{\alpha} \leq 2$,
\item if $\frac{2}{3} \leq \alpha < 1$, the approximation ratio is $\frac{2}{\alpha} \leq 3$,
\item if $\alpha < \frac{2}{3}$, the approximation ratio is $\frac{4}{2-\alpha} < 3$.

\end{enumerate}

\end{theo}
\begin{ourproof}
Case $(i)$ and $(ii)$ follow directly from Lemma \ref{gapratio2}.

We deal with Case $(iii)$. Let us define closed balls centred at $p_{i}$'s as follows: $B_{i} = \{x \in \mathcal{M}:\; \delta(p_{i},x)\leq r_{P} \}$ and $B'_{i} = \{x \in \mathcal{M}:\;\delta(p_{i},x) \leq \alpha r_{P} \}$. We need the following claim.
\begin{ourclaim}\label{claim:lem-new}
 For all $i \in \{2, \, \dots, \, k\}$, $2r_{S_{i}} \geq (2-\alpha) r_{P}$.
\end{ourclaim}

From the proof of Claim~\ref{claim:lem-new}, we have for all $j \in \{ 1, \, \dots, \, k\}$, $|B'_{j}\cap S_{k}| = 1$. Thus, we have $R_{S_{k}} \leq 2\alpha r_{P}$, since $B'_{j}$ cover $\mathcal{M}$.
 Combining this with the fact that $2r_{S_{k}} \geq (2-\alpha) r_{P}$ (Claim~\ref{claim:lem-new}), we have $GR_{S_{k}} \leq \frac{4\alpha}{2-\alpha}$ and consequently
 $$
    \frac{GR_{S_{k}}}{GR_{P}} \leq \frac{4}{2-\alpha} < 3.
 $$
\end{ourproof}

We now prove Claim~\ref{claim:lem-new}.

\begin{ourproof}[of Claim~\ref{claim:lem-new}]
Note that $B_{j}'$'s cover whole of $P$. The case of $i = 2$ follows from the fact that $2r_{S_2}=diam\left(\cal M \right)$. Assume the result is true for some $i \geq 2$. We will show it is true for $S_{i+1}$, if $i \leq k-1$, by contradiction. Suppose $q_{i+1}$ falls into a ball $B'_{j}$ that contains $q_{t}$, for some $t \leq i$. This would imply, $$2 r_{S_{i+1}} \leq \delta(q_{t}, q_{i+1}) \leq 2 \alpha r_{P}$$ Note that as $\alpha < 2/3$, we have $2 \alpha r_{P} < (2-\alpha) r_{P}$. But since, $i\leq k-1$, there exists $p_{t'}$ such that $B'_{t'}$ is empty. That implies we could have selected $p_{t'}$, instead of $q_{i+1}$, to get $$2r_{S_{i+1}} = \min\{ 2r_{S_{i}}, \delta(p_{t'}, S_{i})\} \geq (2-\alpha) r_{P}$$ Note that the last inequality follows from the fact that $2r_{S_{i}} \geq (2-\alpha)r_{P}$ (by induction) and $\delta(p_{t'}, S_{i}) \geq (2-\alpha) r_{P}$.

 Now that we know $q_{i+1}$ falls into a separate ball $B'_{j}$, it is easy to see that $$2r_{S_{i+1}} \geq \min\{ 2r_{S_{i}}, \delta(p_{j}, S_{i})\} \geq (2-\alpha) r_{P}$$
\end{ourproof}

From the results in section~\ref{ssec:lowerboundc}, we have the following corollary to Theorem~\ref{thm-FPI}.
\begin{corr}
The approximation algorithm gives an approximation ratio of

\begin{enumerate}[(i)]
\item $2$ when the metric space is continuous, compact and path connected,
\item $\rho \left(k \right)$, when the metric space is restricted to a unit square in the Euclidean plane, where $\rho \left( k\right)=\frac{\sqrt[4]{27}\sqrt{k}}{\sqrt[4]{3} \sqrt{k} - \sqrt{2}}= \sqrt{3}+O\left(\frac{1}{\sqrt{k}}\right)$, and
\item $3$ when the metric space is restricted to graph metric space.

\end{enumerate}
\end{corr}

\begin{remark}[Time complexity]
\begin{enumerate}
    \item
        In the initialization step of Algorithm~\ref{algo:fpi}, we need to find two points
        $q_{1}, \, q_{2} \in \mathcal{M}$ such that $\delta(q_{1}, q_{2}) = diam({\mathcal{M}})$.
        Given the distance matrix for the metric space $(\mathcal{M}, \delta)$,
        points $q_{1}$ and $q_{2}$ can be computed in time complexity
        $O(n^{2})$. When the points come from Euclidean space $\mathbb{R}^{d}$ and
        $\delta = L_{2}$-norm one can do better. For the case of $d=2, \, 3$
        Ramos~\cite{DBLP:journals/dcg/Ramos01}
        showed that $q_{1}, \, q_{2}$ can be computed in time complexity $O(n \log n)$.
        For the general case of points in $\mathbb{R}^{2}$ and $\delta = L_{2}$-norm,
        Chan~\cite{Chan200620}
        showed that one can get {\em $\epsilon$-approximation} of the diameter
        of the point set in time complexity $O\left( n + \frac{1}{\epsilon^{d- 3/2}}\right)$.

    \item
        Given the distance matrix for $(\mathcal{M}, \delta)$,
        Gonzalez's~\cite{gonzalez-clustering} farthest point insertion method
        can be implemented in $O(nk)$ time complexity.
        Feder and Greene~\cite{DBLP:conf/stoc/FederG88} showed
        that when the points come from Euclidean space ${\mathbb{R}}^{d}$, where $d$ is a constant,
        and $\delta = L_{p}$-norm, then the farthest point procedure
        can be implemented in time complexity $O\left(n\log k \right)$.

\end{enumerate}
\end{remark}

\subsection{Existence of coresets for Euclidean space}
\label{sssec:ealgo}
Next we discuss a $\left( 1+\epsilon \right)$- approximation algorithm when the space $\cal M$ is a set of $n$ points in the Euclidean space. Our approach here to obtain a coreset, which we define below. We need some notations also.

For a metric space $\left( {\cal M} , \delta \right)$ and a set $P\subset \cal M$ we denote the gap ratio of $P$ in $\cal M$ by $GR_{P}^{\cal M}$ and the maximum gap of $P$ in $\cal M$ by $R_{P}^{\cal M}$.

\begin{define}
A subspace $\cal C$ of $\cal M$, is called a $\left( k, \epsilon \right)$-coreset of $\cal M$ if for every set $P\subset \cal C$ of $k$ points we have $GR_{P}^{\cal M} \leq \left( 1+\epsilon \right) GR_{P}^{\cal C}$.
\end{define}

Note that the minimum gap is a property only of the set $P$ and the metric of $\cal M$. Thus, $GR_{P}^{\cal M} \leq \left( 1+\epsilon \right) GR_{P}^{\cal C}$ is equivalent to $R_{P}^{\cal M} \leq \left( 1+\epsilon \right) R_{P}^{\cal C}$.

We now proceed towards the algorithm.

\subsubsection{Static points}
Algorithm~\ref{algo:approx} takes a set of points in the Euclidean space as the input and returns a subset of it which has a gap ratio within a $\left( 1+\epsilon \right)$-factor of the optimal gap ratio.

 Lets assume
 $$
    \epsilon < \frac{1}{2},\;\; \epsilon_1 = \frac{\epsilon}{3+2\epsilon}\;\; \mbox{and}
    \;\; \epsilon_2 = \frac{\epsilon_1 R_{P_1}}{2 \sqrt{d}},
 $$
 where $P_1$ is the output of Algorithm~\ref{algo:approx}.

For analysing the algorithm we first define the following quantities
$$R_{OPT} \coloneqq \min_{\substack{P\subset \mathcal{M},\\ \vert P \vert =k}} \; \max_{q \in {\cal M}} \delta(q,P)
\;\;
 \mbox{and}
 \; \; r_{OPT} \coloneqq\max_{\substack{P\subset \mathcal{M}, \\ \vert P \vert =k}} \;
 \min_{\substack{p,q \in {P},\\ p \neq q}} \frac{\delta(p,q)}{2}.
 $$

\begin{algorithm}
  \caption{Pseudocode of {\em $\left(1 + \epsilon\right)$-Algorithm}}
  \begin{algorithmic}[1]\label{algo:approx}
    \STATE{\bf Input:} Set $\mathcal{M}$ of $n$ points and $k$;
    \STATE  $P_1 \gets Farthest$-$point$-$insertion(\mathcal{M},k) $
    \STATE{\bf Initialize:} Grid $\cal G$ with each cell having side length $\epsilon_2$ and ${\cal C} \gets \left\lbrace \right\rbrace$;
    \FOR{cell $G$ in $\cal G$}
        \STATE $q_G$ is one point randomly chosen from $G \cap \cal M$;
        \STATE ${\cal C} \gets {\cal C} \cup \left\lbrace q_G \right\rbrace$;
    \ENDFOR
    \FOR{$M \subset {\cal C}$ such that $|M|=k$}
    	    \IF {$GR_{M}<GR_P$ or $P$ is not defined}
    	          \STATE $P \gets M$
    	    \ENDIF
    	\ENDFOR
    \STATE{\bf Output:} $P$ and $GR_{P} = \frac{R_P}{r_P}$;
  \end{algorithmic}
\end{algorithm}


To see that $\cal C$ is a coreset we must note that
\begin{eqnarray*}
R_{P}^{\cal M}  \leq  R_{P}^{\cal C} + \sqrt{d} \epsilon_2  = R_{P}^{\cal C} + \epsilon_1 R_{P_1}/2
                \leq  R_{P}^{\cal C} + \epsilon_1 R_{OPT}   \leq R_{P}^{\cal C} + \epsilon_1 R_{P}^{\cal M}
\end{eqnarray*}
i.e., $\left( 1-\epsilon_1 \right) R_{P}^{\cal M} \leq R_{P}^{\cal C}$ or
$$
    R_{P}^{\cal M} \leq \left( 1+\epsilon' \right) R_{P}^{\cal C} \;\; \mbox{where} \;\;
    \epsilon' = \frac{\epsilon_1}{1-\epsilon_1}.
$$

We try to bound the time complexity by estimating the number of grid cells needed to cover $\cal M$.
\begin{lem}\label{lem:noofpoints}
In Algorithm 2, at most $N \coloneqq O(k \lceil \frac{1}{\epsilon_1} \rceil ^d )$ cells cover $\cal M$.
\end{lem}
\begin{ourproof}
Consider, $P_{cov} \subset \cal M$ of size $k$, such that $R_{OPT} = \max_{q \in {\cal M}} \delta \left(q,P_{cov}\right)$. Now, we know that balls of radius $R_{OPT}$ around the points of $P_{cov}$ cover $\cal M$. Each of these balls intersect $O( \lceil \frac{2R_{OPT}}{\epsilon_2} \rceil ^d ) = O( \lceil \frac{1}{\epsilon_1} \rceil ^d )$ grid cells. Thus, $N \coloneqq O(k \lceil \frac{1}{\epsilon_1} \rceil ^d )$ cells cover $\cal M$.
\end{ourproof}
The above lemma shows that the time complexity of a
brute force calculation of gap ratio over $S$ will be
$$
    O\left( N^k \left( k\log k + \left( n-k \right) k \right) \right),
$$
where $O(k\log k)$ is required to compute $r$ and $O( k(n-k) )$ is required to compute $R$ in each iteration; all other steps in Algorithm 2 are polynomial in $n$ and $k$. Note that the time is not polynomial in $k$. We are now ready to prove the main theorem for this section.
\begin{theo}
In Algorithm 2 we have, $GR_P \leq \left(1+\epsilon\right)\cdot GR_{OPT}$.
\end{theo}
\begin{ourproof}
Consider the set $P^*$ of $k$ points in $\cal M$, which gives the minimum gap ratio, $\alpha$, in $\cal M$. Let $r \coloneqq r_{P^*}$. We have $R_{P_1}\leq 2 R_{OPT}$ from~\cite{gonzalez-clustering}. For each $p_i$ in $P^*$, there exists a point $q_i$ in $S$, such that $\delta \left(q_i,p_i\right) \leq \sqrt{d} \epsilon_2$, because $\sqrt{d} \epsilon_2$ is the diameter of each grid cell. From the definition of $\epsilon_2$, we have  $$\delta\left(q_i,p_i\right) \leq \frac{\epsilon_1 R_{P_1}}{2} \leq \epsilon_1 R_{OPT} \leq \epsilon_1 R_{P^*}=\epsilon_1 \alpha r$$ Also note that $\alpha \leq 2$, as the farthest point method itself will yield gap ratio at most $2$. Thus, we have $\delta \left(q_i,p_i\right) \leq r$ (as $\epsilon_1 < \frac{1}{2}$), i.e., $i \neq j$ implies $q_i \neq q_j$. Let $P_2 \coloneqq \left\lbrace q_1, q_2, \ldots, q_k \right\rbrace$ be a set of such $k$ distinct points in $S$. Let us compute the gap ratio of $P_2$. Triangle inequality gives us $R_{P_2} \leq \left( 1+\epsilon_1 \right) \alpha r$ and $r_{P_2} \geq \left( 1-\epsilon_1 \alpha \right) r$. Then the gap ratio of $P_2$ is at most
$$
    \frac{\left( 1+\epsilon_1 \right) \alpha}{\left( 1-\epsilon_1 \alpha \right)}
    \leq \frac{\left( 1+\epsilon_1 \right) \alpha}{\left( 1-2\epsilon_1 \right)}=\left( 1+\epsilon \right) \alpha .
$$
Also by definition, the gap ratio of $P$ is less than the gap ratio of $P_2$.
Thus, we have that gap ratio of $P$ in $S$ is at most $\left( 1+\epsilon \right) \alpha$.
\end{ourproof}

\subsubsection{Streaming}
In a \emph{one-pass streaming model}~\cite{Har-PeledM04, AgarwalHV04, Alon:1996, Muthukrishnan}, that we consider, the data can be read only once and the data is read only, the volume of the data is hude compared to the memory of the storage and typically a sublinear sized sketch of the data is stored. The uniform sample that is to be drawn from the data is drawn from the sketch. In our case, the coreset serves as the sketch.

Algorithm~\ref{algo:approx} works mainly because of the fact that the covering radius is bounded within a constant factor of $R_{OPT}$ by the farthest point insertion algorithm. In the streaming case, the doubling algorithm~\cite{Charikar} for $k$-centre has the same property. We propose Algorithm~\ref{algo:approxs} on the basis of the doubling algorithm. We assume
$$
    \epsilon < \frac{1}{8},\;\; \epsilon_1 = \frac{\epsilon}{2+\epsilon},\;\; \epsilon_2 = \frac{\epsilon_3 R_T}{2 \sqrt{d}}
    \;\; \mbox{where}, \;\; \epsilon_3 = \frac{\epsilon_1}{4 (3+2\epsilon_1)}.
$$
\begin{algorithm}[h]

  \caption{Pseudocode of {\em $\left(1 + \epsilon\right)$-Algorithm} (Streaming case)}
  \begin{algorithmic}[1]\label{algo:approxs}
    \STATE{\bf Input:} Set $\mathcal{M}$ of $n$ points and $k$;
    \STATE {$T \gets \{$first $k$ distinct points$\}$}
    \STATE {$R \gets $smallest interpoint distance in $T$}
     \STATE{\bf Initialize:} Grid $\cal G$ with each cell having side length $\epsilon_2$ and $S \gets \left\lbrace \right\rbrace$;
    \FOR{cell $G$ in $\cal G$}
        \STATE $q_G$ is one point randomly chosen from $G \cap \cal M$;
        \STATE $S \gets S \cup \left\lbrace q_G \right\rbrace$;
    \ENDFOR
    \REPEAT
	    \WHILE{ $\left\vert T \right\vert \leq k$}
    	    \STATE {Get new point $x$;} \label{(A)}
        	\IF {$\delta \left( x,T \right) >2R $}
        		\STATE {$T \gets T \cup \left\lbrace x \right\rbrace$;}
        	\ENDIF

    	\ENDWHILE
    \STATE {$T' \gets \{ \}$ and $S' \gets \{ \}$;}
    \WHILE{$\exists z \in T$ such that $\delta \left( z,T' \right) >2R$}
		\STATE $T' \gets T' \cup \left\lbrace z \right\rbrace$;
		\STATE $R \gets 2R$;
	\ENDWHILE
    \STATE Update grid side-length accordingly (merge neighbouring pairs of columns and rows)
        \FOR{cell $G$ in $\cal G$}
        \STATE $q_G$ is one point randomly chosen from $G \cap \cal M$;
        \STATE $S' \gets S' \cup \left\lbrace q_G \right\rbrace$;
    \ENDFOR
    \STATE $S \gets S'$
	\UNTIL {forever}
  \end{algorithmic}
\end{algorithm}

At Step~\ref{(A)} of Algorithm~\ref{algo:approxs} side-length of the grid cells is always $\epsilon_2$ and $R_{T}~\leq~8R_{OPT}$~\cite{Charikar}. Thus, by the same arguments as in Lemma~\ref{lem:noofpoints}, at most $N \coloneqq O(k \lceil \frac{1}{\epsilon_1} \rceil ^d )$ cells cover $\cal M$ at Step~\ref{(A)} of Algorithm~ \ref{algo:approxs} as well.

To see that $S$ (from Algorithm~\ref{algo:approxs}) is a coreset note that for any subset $V$ of $S$

\begin{eqnarray*}
 R_{V}^{\cal M}  \leq R_{V}^{S} + \frac{\epsilon_3 R_{T}^{\cal M}}{2} \leq R_{V}^{S} + 4 \epsilon_3 R_{OPT}
\end{eqnarray*}

Thus we have,
\begin{eqnarray*}
R_{V}^{{\cal M}}-\epsilon_{1}R_{V}^{{\cal M}}  \leq  R_{V}^{{\cal M}}-4\epsilon_{3}R_{V}^{{\cal M}}
  \leq  R_{V}^{{\cal M}}-4\epsilon_{3}R_{OPT}
  \leq  R_{V}^{S}
\end{eqnarray*}
This implies $\left(1-\epsilon_{1}\right)GR_{V}^{{\cal M}}\leq GR_{V}^{S}$, and therefore
\begin{eqnarray}\label{eq:3}
    GR_{V}^{{\cal M}} \leq \left( 1+\epsilon' \right) GR_{V}^{S} \;\;
    \mbox{where} \;\; \epsilon' = \frac{\epsilon_1}{\left( 1-\epsilon_1\right)}.
\end{eqnarray}
\begin{theo}
Let $S$ be a set obtained from Algorithm~\ref{algo:approxs} and $Q\subset S$ be the set in $S$ with the least gap ratio we have, $GR_Q \leq \left(1+\epsilon\right)\cdot GR_{OPT}$.
\end{theo}
\begin{ourproof}
Consider the set $P^*$ of $k$ points in $\cal M$, which gives the minimum gap ratio, $\alpha$, in $\cal M$. Let $r \coloneqq r_{P^*}$. We have $R_{T}\leq 8 R_{OPT}$ from~\cite{Charikar}. For each $p_i$ in $P^*$, there exists a point $q_i$ in $S$, such that $$\delta \left(q_i,p_i\right) \leq \sqrt{d} \epsilon_2$$ because $\sqrt{d} \epsilon_2$ is the diameter of each grid cell. From the definition of $\epsilon_2$, we have  $\delta\left(q_i,p_i\right) \leq \frac{\epsilon_3 R_{T}}{2} \leq 4 \epsilon_3 R_{OPT} \leq 4 \epsilon_3 R_{P^*}=4 \epsilon_3 \alpha r$. Also note that $\alpha \leq 2$, as the farthest point method itself will yield gap ratio at most $2$. Thus, we have $\delta \left(q_i,p_i\right) \leq r$ (as $\epsilon_3 < \frac{1}{8}$), i.e., $i \neq j$ implies $q_i \neq q_j$.
Let $Q \coloneqq \left\lbrace q_1, q_2, \ldots, q_k \right\rbrace$ be a set of such $k$ distinct points in $S$.

\remove{and
\begin{eqnarray*}
 & R_{V}^{S}\leq R_{V}^{{\cal M}} & \leq\left(1+\epsilon_{1}\right)R_{V}^{{\cal M}}
\end{eqnarray*}

\begin{eqnarray*}
\implies & GR_{V}^{S}\leq\left(1+\epsilon_{1}\right)GR_{V}^{{\cal M}}
\end{eqnarray*}}

Let us compute the gap ratio of $Q$ with respect to $S$. Triangle inequality gives us $R_{Q}^{\cal S} \leq \left( 1+4\epsilon_3 \right) \alpha r$ and $r_{q} \geq \left( 1-4 \epsilon_3 \alpha \right) r$. Then, $$GR_{Q}^{S} \leq \frac{\left( 1+4 \epsilon_3 \right) \alpha}{\left( 1-4 \epsilon_3 \alpha \right)} \leq \frac{\left( 1+\epsilon_3 \right) \alpha}{\left( 1-8\epsilon_1 \right)}=\left( 1+\epsilon_1 \right) \alpha$$ From Equation~\ref{eq:3} we have $$\left( 1- \epsilon_1 \right) GR_{Q}^{\cal M} \leq \left( 1+ \epsilon_1 \right) \alpha$$ which implies $$GR_{Q}^{\cal M} \leq \left( 1+\epsilon\right) \alpha$$

\end{ourproof}

\section{Conclusion}\label{sec:conclusion}

In this work, we generalize the definition of gap ratio given by
Teramoto et al.~\cite{TeramotoAKD06}, for general metric spaces. We show non-existence of a
general lower bound. On the other side, we show constant lower bounds for gap ratio for
connected undirected graphs and metric space of unit squares in the Euclidean plane. We also
show that the problem is NP-hard for discrete and continuous metric spaces. We also design
relevant approximation algorithms and show existence of coresets for the Euclidean space.
Our solutions show connections of picking uniform samples with clustering, packing and
covering. The tightness of the lower bound for unit square is still an open question. Also
the problem of sampling uniformly from a set of points in motion is also an interesting question.


\section*{Acknowledgements}

The authors want to thank Tetsuo Asano and Geevarghese Philip. The authors would also like to thank Dr. Eckard Specht for providing Figure~\ref{fig:pacex}.


\bibliographystyle{alpha}
\bibliography{gap-ratio}

\newcommand{\etalchar}[1]{$^{#1}$}
\begin{thebibliography}{dBCvKO08}

\bibitem[AKOT02]{AsanoKOT02}
T.~Asano, N.~Katoh, K.~Obokata, and T.~Tokuyama.
\newblock {Combinatorial and Geometric Problems Related to Digital Halftoning}.
\newblock In {\em Theoretical Foundations of Computer Vision}, pages 58--71,
  2002.

\bibitem[AKOT03]{AsanoKOT03}
T.~Asano, N.~Katoh, K.~Obokata, and T.~Tokuyama.
\newblock {Matrix Rounding under the $\mbox{L}_p$-Discrepancy Measure and Its
  Application to Digital Halftoning}.
\newblock {\em SIAM Journal on Computing}, 32(6):1423--1435, 2003.

\bibitem[AMS96]{Alon:1996}
N.~Alon, Y.~Matias, and M.~Szegedy.
\newblock {The Space Complexity of Approximating the Frequency Moments}.
\newblock In {\em Proc. 28th Annual ACM Symposium on Theory of Computing},
  pages 20--29, 1996.

\bibitem[Asa06]{Asano06}
T.~Asano.
\newblock {Computational Geometric and Combinatorial Approaches to Digital
  Halftoning}.
\newblock In {\em Theory of Computing 2006, Proceedings of the Twelfth
  Computing: The Australasian Theory Symposium (CATS2006)}, pages 16--19, 2006.

\bibitem[Asa08]{ipl-Asano08}
T.~Asano.
\newblock {Online uniformity of integer points on a line}.
\newblock {\em Information Processing Letters}, 109(1):57--60, 2008.

\bibitem[ASHV04]{AgarwalHV04}
P.~K. Agarwal, S.~Sariel~Har{-}Peled, and K.~R. Varadarajan.
\newblock Approximating extent measures of points.
\newblock {\em Journal of the {ACM}}, 51(4):606--635, 2004.

\bibitem[AT07]{asano-teramoto}
T.~Asano and S.~Teramoto.
\newblock On-line uniformity of points.
\newblock In {\em Book of Abstracts for 8th Hellenic-European Conference on
  Computer Mathematics and its Applications}, pages 21--22, 2007.

\bibitem[BBHS96]{Bange19961}
D.~W. Bange, A.~E. Barkauskas, L.~H. Host, and P.~J. Slater.
\newblock Generalized domination and efficient domination in graphs.
\newblock {\em Discrete Mathematics}, 159(1–3):1 -- 11, 1996.

\bibitem[CCFM97]{Charikar}
M.~Charikar, C.~Chekuri, T.~Feder, and R.~Motwani.
\newblock {Incremental Clustering and Dynamic Information Retrieval}.
\newblock In {\em Proc. 29th Annual ACM Symposium on Theory of Computing},
  pages 626--635, 1997.

\bibitem[CCL96]{ChainChin1996147}
Y.~Chain-Chin and R.~C.~T. Lee.
\newblock The weighted perfect domination problem and its variants.
\newblock {\em Discrete Applied Mathematics}, 66(2):147 -- 160, 1996.

\bibitem[CDS12]{Cheng}
S.-W. Cheng, T.~K. Dey, and J.~Shewchuk.
\newblock {\em {Delaunay Mesh Generation}}.
\newblock Chapman \& Hall/CRC, 1st edition, 2012.

\bibitem[Cha01]{book-chazelle}
B.~Chazelle.
\newblock {\em {The Discrepancy Method - Randomness and Complexity}}.
\newblock Cambridge University Press, 2001.

\bibitem[Cha06]{Chan200620}
T.~M. Chan.
\newblock Faster core-set constructions and data-stream algorithms in fixed
  dimensions.
\newblock {\em Computational {G}eometry}, 35(1–2):20 -- 35, 2006.

\bibitem[CS03]{CollinsS03}
C.~R. Collins and K.~Stephenson.
\newblock A circle packing algorithm.
\newblock {\em Computational Geometry}, 25(3):233 -- 256, 2003.

\bibitem[dBCvKO08]{Berg}
M.~de~Berg, O.~Cheong, M.~van Kreveld, and M.~Overmars.
\newblock {\em {Computational Geometry: Algorithms and Applications}}.
\newblock Springer, 3rd ed. edition, 2008.

\bibitem[DG94]{Dobkin}
D.~P. Dobkin and D.~Gunopulos.
\newblock {Computing the Rectangle Discrepancy}.
\newblock In {\em Proc. 10th Annual Symposium on Computational Geometry}, pages
  385--386, 1994.

\bibitem[Doe14]{Doerr}
B.~Doerr.
\newblock A lower bound for the discrepancy of a random point set.
\newblock {\em Journal of Complexity}, 30(1):16 -- 20, 2014.

\bibitem[FG88]{DBLP:conf/stoc/FederG88}
T.~Feder and D.~H. Greene.
\newblock {Optimal Algorithms for Approximate Clustering}.
\newblock In {\em Proc. 20th Annual {ACM} Symposium on Theory of Computing},
  pages 434--444, 1988.

\bibitem[FKP05]{FialaKP}
J.~Fiala, J.~Kratochv\'{\i}l, and A.~Proskurowski.
\newblock Systems of distant representatives.
\newblock {\em Discrete Applied Mathematics}, 145(2):306--316, 2005.

\bibitem[GJ90]{Garey:1990:CIG:574848}
M.~R. Garey and D.~S. Johnson.
\newblock {\em {Computers and Intractability; A Guide to the Theory of
  NP-Completeness}}.
\newblock W. H. Freeman \& Co., New York, NY, USA, 1990.

\bibitem[Gon85]{gonzalez-clustering}
T.~F. Gonzalez.
\newblock Clustering to minimize the maximum intercluster distance.
\newblock {\em Theoretical Computer Science}, 38:293 -- 306, 1985.

\bibitem[HM04]{Har-PeledM04}
Sariel Har{-}Peled and Soham Mazumdar.
\newblock On coresets for k-means and k-median clustering.
\newblock In {\em Proc. 36th Annual {ACM} Symposium on Theory of Computing},
  pages 291--300, 2004.

\bibitem[HM11]{Hinrichs}
A.~Hinrichs and L.~Markhasin.
\newblock On lower bounds for the -discrepancy.
\newblock {\em Journal of Complexity}, 27(2):127 -- 132, 2011.

\bibitem[KN74]{Kuipers}
L.~Kuipers and H.~Niederreiter.
\newblock {\em {Uniform distribution of sequences}}.
\newblock Wiley-Interscience [John Wiley \& Sons], New York, 1974.
\newblock Pure and Applied Mathematics.

\bibitem[Kra94]{Kratochvl1994233}
J.~Kratochv\'{i}l.
\newblock A special planar satisfiability problem and a consequence of its
  np-completeness.
\newblock {\em Discrete Applied Mathematics}, 52(3):233 -- 252, 1994.

\bibitem[Kup87]{kuperberg}
W.~Kuperberg.
\newblock An inequality linking packing and covering densities of plane convex
  bodies.
\newblock {\em Geometriae Dedicata}, 23(1):59--66, 1987.

\bibitem[LR02]{Locatelli}
M.~Locatelli and U.~Raber.
\newblock Packing equal circles in a square: a deterministic global
  optimization approach.
\newblock {\em Discrete Applied Mathematics}, 122(1–3):139 -- 166, 2002.

\bibitem[Mat99]{book-matousek-1999}
J.~Matou\v{s}ek.
\newblock {\em {Geometric Discrepancy: An Illustrated Guide}}.
\newblock Algorithms and Combinatorics. Springer, 1999.

\bibitem[Mut05]{Muthukrishnan}
S.~Muthukrishnan.
\newblock {Data Streams: Algorithms and Applications}.
\newblock {\em Foundations and Trends in Theoretical Computer Science},
  1(2):117--236, 2005.

\bibitem[N{\"O}97]{Nurmela-dcg-O97}
K.~J. Nurmela and P.~R.~J. {\"O}sterg{\aa}rd.
\newblock {Packing up to 50 Equal Circles in a Square}.
\newblock {\em Discrete {\&} Computational Geometry}, 18(1):111--120, 1997.

\bibitem[N{\"O}99]{Nurmela-dcg-O99}
K.~J. Nurmela and P.~R.~J. {\"O}sterg{\aa}rd.
\newblock {More Optimal Packings of Equal Circles in a Square}.
\newblock {\em Discrete {\&} Computational Geometry}, 22(3):439--457, 1999.

\bibitem[N{\"O}adS99]{Nurmela-cmb-O99}
K.~J. Nurmela, P.~R.~J. {\"O}sterg{\aa}rd, and R.~aus~dem Spring.
\newblock {Asymptotic Behavior of Optimal Circle Packings in a Square}.
\newblock {\em Canadian Mathematical Bulletin}, 42(3):380--385, 1999.

\bibitem[OKO12]{Ong}
M.~S. Ong, Y.~C. Kuang, and M.~P.-L. Ooi.
\newblock Statistical measures of two dimensional point set uniformity.
\newblock {\em Computational Statistics {\&} Data Analysis}, 56(6):2159 --
  2181, 2012.

\bibitem[Ram01]{DBLP:journals/dcg/Ramos01}
E.~A. Ramos.
\newblock {An Optimal Deterministic Algorithm for Computing the Diameter of a
  Three-Dimensional Point Set}.
\newblock {\em Discrete {\&} Computational Geometry}, 26(2):233--244, 2001.

\bibitem[RBGP06]{Romero}
V.~J. Romero, J.~V. Burkardt, M.~D. Gunzburger, and J.~S. Peterson.
\newblock {Comparison of pure and “Latinized” centroidal Voronoi
  tessellation against various other statistical sampling methods}.
\newblock {\em Reliability Engineering {\&} System Safety}, 91(10–11):1266 --
  1280, 2006.

\bibitem[STCT02]{SadakaneTT02}
K.~Sadakane, N.~Takki-Chebihi, and T.~Tokuyama.
\newblock {Discrepancy-Based Digital Halftoning: Automatic Evaluation and
  Optimization}.
\newblock In {\em Theoretical Foundations of Computer Vision}, pages 301--319,
  2002.

\bibitem[T\'83]{toth}
G.~F. T\'{o}th.
\newblock {New Results in the Theory of Packing and Covering}.
\newblock In P.~M. Gruber and J.~M. Wills, editors, {\em {Convexity and Its
  Applications}}, pages 318--359. Birkhäuser Basel, 1983.

\bibitem[TAKD06]{TeramotoAKD06}
S.~Teramoto, T.~Asano, N.~Katoh, and B.~Doerr.
\newblock {Inserting Points Uniformly at Every Instance}.
\newblock {\em IEICE (Institute Electronics, Information and Communication
  Engineers) TRANSACTIONS on Information and Systems}, 89-D(8):2348--2356,
  2006.

\bibitem[ZCC{\etalchar{+}}11]{ipl-ZhangCCTT11}
Y.~Zhang, Z.~Chang, F.~Y.~L. Chin, H.-F. Ting, and Y.~H. Yung H.~Tsin.
\newblock Uniformly inserting points on square grid.
\newblock {\em Information Processing Letters}, 111(16):773--779, 2011.

\end{thebibliography}

\end{document}